\documentclass{article}

\usepackage{arxiv}

\usepackage[bigfiles]{pdfbase}
\ExplSyntaxOn
\NewDocumentCommand\embedvideo{smm}{
  \group_begin:
  \leavevmode
  \tl_if_exist:cTF{file_\file_mdfive_hash:n{#3}}{
    \tl_set_eq:Nc\video{file_\file_mdfive_hash:n{#3}}
  }{
    \IfFileExists{#3}{}{\GenericError{}{File~`#3'~not~found}{}{}}
    \pbs_pdfobj:nnn{}{fstream}{{}{#3}}
    \pbs_pdfobj:nnn{}{dict}{
      /Type/Filespec/F~(#3)/UF~(#3)
      /EF~<</F~\pbs_pdflastobj:>>
    }
    \tl_set:Nx\video{\pbs_pdflastobj:}
    \tl_gset_eq:cN{file_\file_mdfive_hash:n{#3}}\video
  }
  \pbs_pdfobj:nnn{}{dict}{
    /Type/RichMediaInstance/Subtype/Video
    /Asset~\video
    /Params~<</FlashVars (
      source=#3&
      skin=SkinOverAllNoFullNoCaption.swf&
      skinAutoHide=true&
      skinBackgroundColor=0x5F5F5F&
      skinBackgroundAlpha=0.75
    )>>
  }
  \pbs_pdfobj:nnn{}{dict}{
    /Type/RichMediaConfiguration/Subtype/Video
    /Instances~[\pbs_pdflastobj:]
  }
  \pbs_pdfobj:nnn{}{dict}{
    /Type/RichMediaContent
    /Assets~<<
      /Names~[(#3)~\video]
    >>
    /Configurations~[\pbs_pdflastobj:]
  }
  \tl_set:Nx\rmcontent{\pbs_pdflastobj:}
  \pbs_pdfobj:nnn{}{dict}{
    /Activation~<<
      /Condition/\IfBooleanTF{#1}{PV}{XA}
      /Presentation~<</Style/Embedded>>
    >>
    /Deactivation~<</Condition/PI>>
  }
  \hbox_set:Nn\l_tmpa_box{#2}
  \tl_set:Nx\l_box_wd_tl{\dim_use:N\box_wd:N\l_tmpa_box}
  \tl_set:Nx\l_box_ht_tl{\dim_use:N\box_ht:N\l_tmpa_box}
  \tl_set:Nx\l_box_dp_tl{\dim_use:N\box_dp:N\l_tmpa_box}
  \pbs_pdfxform:nnnnn{1}{1}{}{}{\l_tmpa_box}
  \pbs_pdfannot:nnnn{\l_box_wd_tl}{\l_box_ht_tl}{\l_box_dp_tl}{
    /Subtype/RichMedia
    /BS~<</W~0/S/S>>
    /Contents~(embedded~video~file:#3)
    /NM~(rma:#3)
    /AP~<</N~\pbs_pdflastxform:>>
    /RichMediaSettings~\pbs_pdflastobj:
    /RichMediaContent~\rmcontent
  }
  \phantom{#2}
  \group_end:
}
\ExplSyntaxOff

\usepackage{graphicx} 
\graphicspath{ {images/}} 

\usepackage[authoryear,round]{natbib}

\usepackage{amsmath} 
\usepackage[separate-uncertainty=true]{siunitx}
\usepackage[utf8]{inputenc} 
\usepackage{xcolor}
\usepackage{subcaption} 
\usepackage{float} 
\usepackage{multirow}
\usepackage{array}
\usepackage{adjustbox}
\usepackage{animate}
\usepackage[normalem]{ulem} 
\usepackage{tabularx,booktabs}
\usepackage{amssymb}
\usepackage{media9}
\usepackage{placeins}
 \usepackage{url}

\newcommand\figref[1]{figure \ref{#1}}               

\usepackage{gensymb}

\title{Fast proton transport and neutron production in proton therapy using Fourier neural operators}

\author{
 Francesco Blangiardi \\
  Fraunhofer ENAS, Technical University Chemnitz \\
  Chemnitz, Germany \\
  \texttt{francesco.blangiardi@enas.fraunhofer.de} \\
   \And
 Hunter N. Ratliff \\
  Western Norway University of Applied Sciences \\
  Bergen, Norway \\
  \texttt{Hunter.Nathaniel.Ratliff@hvl.no} \\
  \And
 Fabian Teichert \\
  Fraunhofer ENAS, Technical University Chemnitz\\
  Chemnitz, Germany \\
  \texttt{fabian.teichert@enas.fraunhofer.de} \\
  \And
 Kristian Smeland Ytre-Hauge \\
  University of Bergen \\
  Bergen, Norway \\
  \texttt{kristian.ytre-hauge@uib.no} \\
  \And
 Jan Langer \\
  Fraunhofer ENAS \\
  Chemnitz, Germany \\
  \texttt{jan.langer@enas.fraunhofer.de} \\
  \And
 Ilker Meric \\
  Western Norway University of Applied Sciences \\
  Bergen, Norway \\
  \texttt{Ilker.Meric@hvl.no} \\
}

\begin{document}
\maketitle


\begin{abstract}
\textit{Objective}: Real-time adaptive proton range verification systems based on produced neutrons require accurate information on their non-isotropic momentum distributions within short times, for which Monte Carlo (MC) methods are too computationally expensive. 
We present a surrogate model based on Fourier Neural Operators (FNO) for fast prediction of angle- and energy-resolved proton transport and neutron production within proton therapy.

\textit{Approach}: We treat the irradiated phantom and the proton beam's state as depth-evolving series, respectively of different materials, and of spatial, angular and energy phase space density distributions. 
The task is solved auto-regressively by learning changes in the distributions of protons and those of produced neutrons.
For training and evaluation, two datasets of 47 MC simulations featuring different primary intensities were produced. Simulated geometries were extracted from a thoracic CT scan as series of laterally homogeneous materials.

\textit{Main Results}: An average relative $L^2$ discrepancy of $0.067$ and $0.137$ was achieved by the predicted proton and neutron distributions, respectively.
This corresponded to an average gamma passing rate in the spatial distributions of $99.95\%$ and $99.40\%$. Training with higher primary intensities led to improvements between $12\%$ and $30\%$ in density metrics.
Inference over depths of \SI{40}{cm} at a resolution of \SI{0.5}{mm} required on average \SI{23.17}{s} per beam.

\textit{Significance}: The proposed proton beam surrogate generates accurate spatial and momentum distributions of neutrons at MC-level accuracy within seconds, while demonstrating robust generalization with respect to irradiated geometry and beam characteristics. This approach can be used for prototyping and operation of range verification systems, other tasks such as neutron dose estimation, and can be extended to include other kinds of secondary emissions.

\end{abstract}

\section{Introduction}
Proton Therapy (PT) is a type of cancer treatment employing ionizing radiation that has been steadily rising in adoption over the past decades. The main reason for its adoption is due to the characteristic Bragg Peak in its depth-dose deposition profile, which enables intense irradiation of narrow volumes while sparing nearby healthy tissues.

To uniformly irradiate the entire tumor, the latest techniques include Intensity Modulated Proton Therapy (IMPT) \citep{moreno_intensity_2019, farr_new_2018}, where numerous pencil beams with a narrow spot size are delivered from different directions within a single treatment session.
However, this involves modulating the energy of each pencil beam depending on the stopping power of the traversed materials, and can therefore lead to errors stemming from uncertainties in the stopping power calculations, or from incoherence between the planned and actual paths traversed during treatment due to anatomical changes or patient motion \citep{mohan_proton_2017}.

\begin{figure}[htbp]
	\centering
	\includegraphics[width=0.5\textwidth, trim={0 0 0 0},clip]{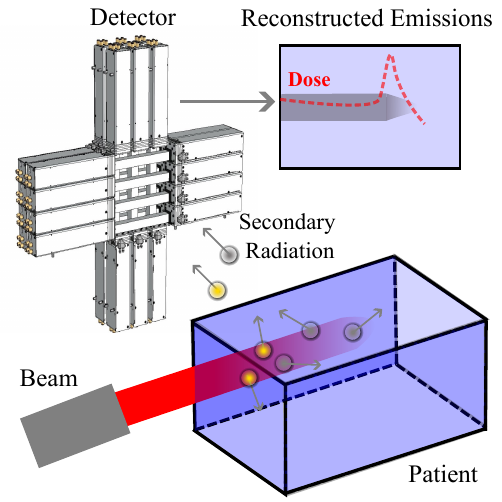}
	\caption{ 
        Schematic of the NOVO range verification system. Detection is performed for both prompt gamma rays and neutrons produced during treatment.
    }
	\label{fig:rv_system}
\end{figure}

To properly assess the effect of these errors, solutions such as \textit{in vivo} range verification systems have been designed to determine the proton range directly during treatment \citep{parodi_vivo_2018}. A family of such systems relies on the detection and characterization of secondary particles produced during the treatment, such as prompt gamma rays, using suitable detectors \citep{perali_prompt_2014, polf_imaging_2015}. The core idea is to reconstruct their emission profile within the patient, which can subsequently be correlated with the beam's range and delivered dose. More recent works are focused on incorporating secondary fast neutrons in a multi-particle approach \citep{lerendegui-marco_simultaneous_2022, Meric_2023, setterdahl_enhancing_2025}, thus obtaining complementary information from different byproducts of the treatment while mitigating issues related to limited counting statistics. 
In the case of the NOVO project \citep{Meric_2023}, to which this work belongs, geometric constraints and detector positioning, coupled with correlations between the proton energy and angular divergence of produced neutrons, can significantly affect the quantity and information content of detected particles. 
In \figref{fig:rv_system}, a schematic of the NOVO range verification system is shown.

Due to the complexity of such systems and factors affecting their response, their development requires extensive prototyping, during which the robustness of the system to range shifts has to be assessed \citep{setterdahl_evaluating_2025, Setterdahl_unets_2024}. Moreover, to detect range shifts during treatment, range verification pipelines may use both the measured detector response and the expected one computed from the treatment plan \citep{gueth_machine_2013}. Although Monte Carlo (MC) codes are considered the gold standard in terms of accuracy to perform such tasks \citep{paganetti_range_2012}, their high computational cost significantly hinders research and implementation in time-constrained clinical workflows such as online adaptive proton therapy, prompting the development of fast approximations of the beam traversal and production of secondary particles.

At a fundamental level, doing so requires calculating the phase space of protons along the beam path, accounting for physical effects such as energy loss, range straggling, Coulomb scattering, and nuclear reactions with the various materials traversed. The relevant physics can be modeled through the Linear Boltzmann Transport Equation (LBTE), but its integro-differential form over the required dimensions makes numerical solutions computationally challenging.
Traditionally, fast proton beam algorithms \citep{hong_pencil_1996, soukup_pencil_2005} for calculating proton transport in PT have primarily been developed to enable fast dose calculations within treatment planning systems, which is achieved analytically by approximating the problem geometry and its high-dimensional physics using the Fermi-Eyges theory \citep{gottschalk_techniques_2012}. Since these methods are normally unsuited in highly heterogeneous scenarios \citep{taylor_pencil_2017}, task-specific AI-based solutions have also been developed to refine their estimations \citep{wu_improving_2021}, or to avoid the explicit computation of the beam's phase space throughout the phantom \citep{neishabouri_dose_2020, pastor-serrano_millisecond_2022}, while still achieving dose predictions with MC-level accuracy. Although similar approaches have been followed to compute emissions of prompt gamma rays \citep{xiao_prompt_2024}, doing so for neutrons requires modeling the correlations between their non-isotropic angular and energy distributions with those of the proton beam \citep{ytre-hauge_monte_2019}. On that front, analytical solutions estimating neutron emissions have been developed to compute their delivered equivalent dose within passively scattered proton therapy \citep{shrestha_stray_2022} as well as IMPT \citep{schneider_neutrons_2017}, but the fast, MC-accurate prediction of joint energy and angular distributions of neutrons in proton therapy through AI methods remains unexplored.

To accurately compute the proton transport and bridge requirements for online-adaptive proton therapy, solutions that relax simplifications of proton beam algorithms and solve the LBTE as a depth-dependent Partial Differential Equation (PDE) through methods borrowed from traditional computational physics have also been developed \citep{burlacu_deterministic_2023}. More recent works following this line of research have focused on efficiently handling fully heterogeneous geometries, speed-resolution trade-offs, and modeling of more challenging phenomena such as catastrophic proton scatters and biological effectiveness \citep{stammer_deterministic_2025, zhang_deterministic_2025, ashby_efficient_2025}.

Recent advancements in Scientific Machine Learning have focused on solving PDEs of evolving systems in a similar form, by using neural networks to approximate the time- or space-dependent evolution of the underlying physical system while aiming for a significant speed-up. 
Among them, Neural Operators \citep{kovachki_neural_2024, azizzadenesheli_neural_2024}, and in particular Fourier Neural Operators (FNO) \citep{li_fourier_2021}, have gained popularity in the modeling of physical systems due to their ability to efficiently perform the operator mapping for the entire target domain at once. 
Among the many use-cases, these techniques have been applied to particle transport and the LBTE to predict photon radiation distributions from positron emission tomography and computed tomography examinations \citep{hu_predicting_2025}.

As, to the best of our knowledge, no other work has applied such methods for fast surrogate modeling of proton therapy, this paper represents a first study under simplified conditions of their application in the prospects of new range verification systems. We use FNOs to propagate the proton phase space distributions through depth within a laterally homogeneous phantom and subsequently use them to predict the phase space of produced neutrons. We specifically target the joint spatial, energy and direction distributions of the neutrons because of their necessity in neutron-based range verification systems, but we believe our approach to be relevant for other kinds of secondary emissions, and for other tasks relevant to proton therapy, such as prediction of neutron-equivalent dose or neutron-induced reactions. 

\section{Materials and Methods}
\label{sec:materials_and_methods}

\subsection{Theoretical background}
We aim to model the proton transport in similar conditions to those simulated within 
MC codes. Such frameworks model the proton energy loss primarily through Coulomb inelastic interactions with electrons (ionization and excitation), often employing condensed-history techniques such as the Continuous Slowing Down Approximation (CSDA) with energy-loss straggling (ELS) and angular straggling approximations to account for many small ionization events. They also include Coulomb elastic scattering with nuclei and discrete catastrophic nuclear reactions, all determined by material-dependent cross-sectional nuclear data (evaluated or physics-model-approximated) representing their probability of occurring per unit path length traversed. Although MC codes also provide a highly accurate modeling of the time dimension, we do not handle it explicitly as it is generally left out of most analytical solutions. Instead, we consider the proton phase space distribution as its time-integrated form, and we target the initial neutron phase space only at production time (i.e. without simulating its transport).

Under these premises, we consider a volume of interest $\mathbf{P} \subset \mathbb{R}^3$ representing the irradiated phantom and associated with a material density function $\rho: \mathbf{P} \rightarrow \mathbb{R}$. We index $\mathbf{P}$ by $\mathbf{r} \equiv (x,y,z) \in \mathbf{P}$, indicating any coordinate system where $z$ is coherent with the beam axis and positive along the phantom's depth. We also denote the energy dimension as $E \in [E_{min}, E_{max}]$, and define $\mathbf{\Omega} \in \mathbb{S}^2$ as the angular directions lying on the unit sphere, which we represent both through the polar and azimuthal angles $\theta \in [0,\pi],\phi \in [0,2\pi)$ and the components of the directional vector $\mu$, $\eta$ and $\xi$ along $x$, $y$ and $z$ respectively. Finally, we denote inelastic Coulomb, elastic Coulomb and catastrophic scattering reaction cross sections respectively as $\sigma_{i,s}(\mathbf{r}, \mathbf{\Omega}, E \rightarrow E^{'})$, $\sigma_{e,s}(\mathbf{r}, \mathbf{\Omega} \rightarrow \mathbf{\Omega^{'}}, E)$ and $\sigma_{c,s}(\mathbf{r}, \mathbf{\Omega} \rightarrow \mathbf{\Omega^{'}}, E \rightarrow E^{'})$.

We then follow the formulation from \citet{zhang_deterministic_2025} and describe the proton phase space density $\psi^{p}(\mathbf{r}, \mathbf{\Omega},E)$, representing the probability distribution of the protons' momentum within the entire phantom, through the LBTE under the CSDA, ELS, and Fokker--Planck approximation for small elastic scattering angles as
\begin{equation}
    \begin{aligned}
    \mu\frac{\partial \psi^p}{\partial x} +\eta\frac{\partial \psi^p}{\partial y}+\xi\frac{\partial \psi^p}{\partial z} &= \frac{\rho}{2}\sigma_{tr}(E)\left[\frac{\partial}{\partial \xi}(1-\xi^2)\frac{\partial}{\partial \xi} + \frac{1}{1-\xi^2}\frac{\partial^2}{\partial \phi^2}\right]\psi^p  \\ 
    &\quad +\rho\frac{\partial}{\partial E}(S(\mathbf{r}, E)\psi^p)+\frac{\rho}{2}\frac{\partial^2}{\partial E^2}(T(\mathbf{r}, E)\psi^p) \\ 
    &\quad+ \int_{4\pi}\int_{E'}^{E_{max}}\sigma_{c,s}\psi^p \mathrm{d}\mathbf{\Omega}\mathrm{d}E - \sigma_{c,t}(\mathbf{r}, E)\psi^p
    \end{aligned}
    \label{eq:bte_full}
\end{equation}
Here, the three rows on the right-hand side of the equation respectively describe the beam broadening due to elastic scatters (with $\sigma_{tr}(E)$ being the rate of angular divergence obtainable from $\sigma_{e,s}$), energy loss due to inelastic scatters (modeled by $S(\mathbf{r}, E)$ and $T(\mathbf{r}, E)$, representing the first two statistical moments of the post-reaction energy governed by $\sigma_{i,s}$), and catastrophic reactions including large-angle scatters and absorption (computed from $\sigma_{c,s}$ and its integral over directions $\sigma_{c,t}$).

Given the characterization of $\psi^{p}$ across the entire phase space domain, it is then possible to find the initial neutron phase space density as:
\begin{equation}
    \psi^{n}(\mathbf{r}, \mathbf{\Omega}^{'} E^{'}) = \int_{4\pi}\int_{E^{'}}^{E_{max}}\sigma^{n}(\mathbf{r}, \mathbf{\Omega} \rightarrow \mathbf{\Omega}^{'}, E \rightarrow E^{'})\psi^{p}(\mathbf{r}, \mathbf{\Omega}, E)\mathrm{d}E\mathrm{d}\mathbf{\Omega} 
\label{eq:density_neutrons}
\end{equation}
where $\sigma^{n}$ represents the energy- and direction-dependent neutron production cross section from proton reactions, describing the distribution of produced neutrons from nuclear reactions induced by the proton traversal. 

\subsection{Proposed framework}
\label{subsec:surrogate_design}

As equations \eqref{eq:bte_full} and \eqref{eq:density_neutrons} present high-dimensional integral and differential terms requiring detailed knowledge of employed cross sections, their numerical solution is quite challenging. In the case of equation \eqref{eq:density_neutrons}, the double-differential secondary particle production cross section data determining $\sigma^{n}$ are not available for all incident proton states, and are usually supplemented with nuclear physics reaction models \citep{INCL46_ref} within MC codes, meaning that a comprehensive MC-based fitting would have to be performed over different materials and proton energies to extract the cross-section data. Within this work, we therefore aim to approximate the solution operators of both equations in a data-driven way using neural networks.

By grouping the mass density $\rho$ and any other material-related quantity involved in the employed cross section under a geometry function $a: \mathbf{P} \rightarrow \mathbb{R}^n$, 
denoting the domain of $\psi^p$ and $\psi^n$  as $\Gamma \equiv (\mathbf{r}, \mathbf{\Omega}, E)$ and the domain of the proton density at the entrance of the phantom as $\Gamma_0 = \{\gamma \in \Gamma : z = 0\}$, the objective is to find two approximated operators of the form
\begin{gather}
    G^{p} : \Psi_{0}^{p} \times \mathcal{A} \rightarrow \Psi^{p} \label{eq:density_op_protons}\\ 
    G^{n} : \Psi^{p} \times \mathcal{A} \rightarrow \Psi^{n} \label{eq:density_op_neutrons} 
\end{gather}
where $\Psi^{p}$ and $\Psi^{n}$ respectively denote the space of all possible proton and neutron density functions over $\Gamma$, $\Psi_{0}^{p}$ represents the space of possible incident proton distributions over $\Gamma_{0}$, while $\mathcal{A}$ denotes the space of possible geometry functions $a$. The described proton operator $G^{p}$ therefore maps the geometry function $a$ and the incident proton density $\psi^{p}_0$ to the entire proton density $\psi^{p}$, while the neutron operator $G^{n}$ will use such result together with $a$ to estimate the neutron phase space density $\psi^{n}$.

As learning the direct mappings represented by equations \eqref{eq:density_op_protons} and \eqref{eq:density_op_neutrons} poses significant challenges for AI implementation and data collection due to the high dimensionality of the problem, we apply some approximations to make the proposed solution feasible. We therefore avoid modeling the transport of back-scattered protons and consider the proton transport as a unidirectional process that evolves sequentially along $z$, which we remove from the phase space domain $\Gamma$ and treat as a discretized dimension representing a pseudo-time step. We can then rewrite operators $G^{p}$ and $G^{n}$ as
\begin{gather}
    G^{p} : \Psi_{k}^{p} \times \mathcal{A}_{k} \rightarrow \Psi_{k+1}^{p} \label{eq:prot_op_zdiscr}\\
    G^{n} : \Psi_{k}^{p} \times \mathcal{A}_{k} \rightarrow \Psi_{k}^{n} \label{eq:neutr_op_zdisc} 
\end{gather}
where $\Psi_{k}^{p}$ generalizes the concept of $\Psi_{0}^{p}$ to a generic domain $\Gamma_k = \{\gamma \in \Gamma : z = z_k\}$, while $\mathcal{A}_{k}$ and $\Psi_{k}^{n}$ respectively represent the space of possible depth-discretized geometry and neutron density functions between surfaces identified by $z_k$ and $z_{k+1}$. Here, each $z_{k}$ associated with step $k \in \{0, \dots ,k_{max}\}$ represents equally-spaced increasing depths.
Under such conditions, the proposed $G^{p}$ and $G^{n}$ operators can therefore be used to recursively map the incident proton density $\psi_{k}^{p}$ and the geometry function $a_{k}$ associated to step $k$ to densities $\psi_{k+1}^{p}$ and $\psi_{k}^{n}$.

In this form, $\psi_{k}^{p}$ and $\psi_{k}^{n}$ no longer represent a probability distribution along the entire extent of $\mathbf{P}$ but only between discretized regions of space, meaning that the absolute number of transported and produced particles among different depths cannot be handled by $G^{p}$ and $G^{n}$.
For such purpose, we define two auxiliary functionals $F^{p} : \Psi_{k}^{p} \times \mathcal{A}_{k} \rightarrow \mathbb{R}$ and $F^{n}:\Psi_{k}^{p} \times \mathcal{A}_{k} \rightarrow \mathbb{R}$ such that:
\begin{gather}
    F^{p}(\psi_{k}^{p}, a_{k}) = I_{k+1}^p \label{eq:prot_int_func}\\
    F^{n}(\psi_{k}^{p}, a_{k}) = I_k^n \label{eq:neutr_int_func} 
\end{gather}
where $I_{k+1}^p$ and $I_k^n$ respectively represent a relative amount of transported protons and produced neutrons after step $k$.
By assuming $\psi_{0}^{p}$ and the primary intensity as known, since they only depend on characteristics of the treatment head and its relative positioning with respect to $\mathbf{P}$, the proposed surrogate can therefore fully reconstruct the phase space densities $\psi^{p}$ and $\psi^{n}$ up to an error introduced by the discretization of $z$.

\begin{figure}[htbp]
	\centering
	\includegraphics[width=1.00\textwidth, trim={0 0 0 0},clip]{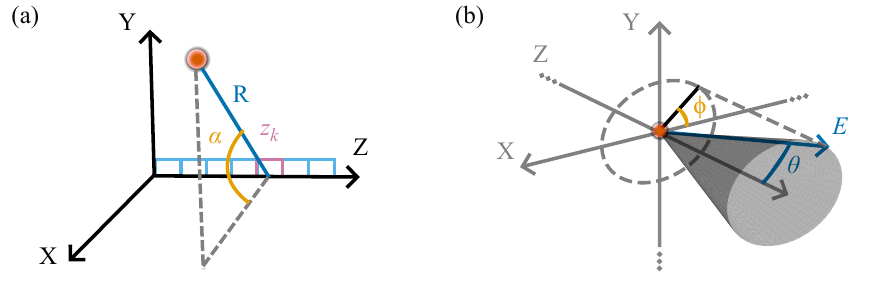}
	\caption{Visualization of the relevant dimensions in the proposed surrogate: (a) the spatial dimensions; (b) the directional and energy dimensions. Retained dimensions ($R$, $E$ and $\theta$) are colored in blue, while those assumed as independently distributed (meaning $\alpha$ and $\phi$) are shown in orange. The discretized depth position $z_{k}$ is instead colored in pink as it is handled only indirectly by the surrogate's operators. 
    }
    \label{fig:dimensional_reduction}
\end{figure}

As further simplifications, within this work we employ uniform phantoms along the $(x, y)$ dimensions, meaning that we limit $\mathcal{A}_{k}$ to represent each function $a_{k}$ as constant, and we consider proton beam spots with radially symmetric shape and directional divergence. Moreover, we consider the spatial dimensions in cylindrical coordinates as $\mathbf{r} \equiv (R, \alpha, z)$, from which independence of the azimuthal angle $\alpha$ with respect to the other dimensions follows because of geometrical symmetries. We also 
approximate interactions between the azimuthal directional angle $\phi$ and other dimensions to be statistically negligible,
which is motivated by cross-section data typically featuring uniform cross sections over the post-interaction azimuthal angle with respect to the incident particle's direction, and by most proton reactions being of small-scattering nature. As such, the domain in which the phase space densities are defined can be reduced to $\Gamma_{k} = (R, E, \theta)$, meaning that other dimensions can be handled outside of the operators. These dimensions are visualized together with other dimensions in \figref{fig:dimensional_reduction}.

\subsection{Data Collection}
\label{subsec:data_collection}

To generate training and testing data representing phase space densities $\psi^{p}$ and $\psi^{n}$, MC radiation transport simulations were conducted. To this end, the PHITS (Particle and Heavy Ion Transport code System) general-purpose Monte Carlo particle transport code, version 3.341 \citep{PHITS333_ref}, was used for modeling proton transport, interactions, and subsequent production of secondary particles, with a particular focus on secondary neutrons. 

\begin{figure}[htbp]
	\centering
	\includegraphics[width=1.00\textwidth, trim={0 0 0 0},clip]{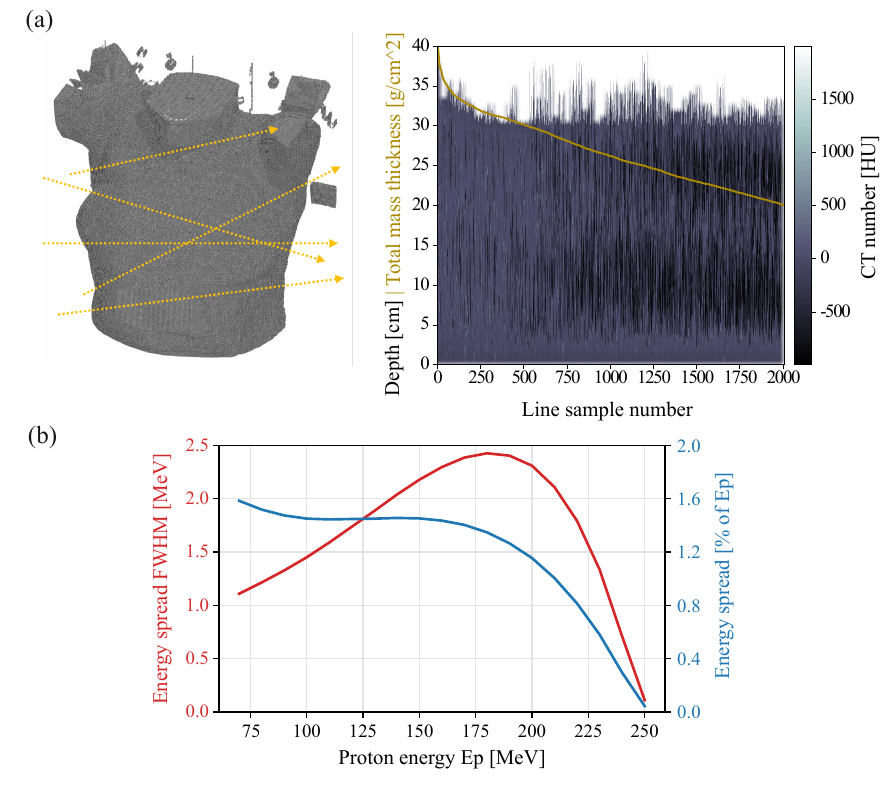}
	\caption{ Details of the performed simulations: (a) the phantom utilized to sample candidate material sequences by casting randomly-directed lines (left) and corresponding material profiles per depth in Houndfield Units (HU) (right); (b) initial energy spread function as a function of the average energy. 
    }
    \label{fig:megafigure}
\end{figure}

The generation of series of realistic material compositions $\{a_{0}, \dots, a_{k_{max}}\}$, all parametrized in terms of material composition and density, was performed using a DICOM file from the patient dataset of the Lung CT Segmentation Challenge, which is publicly available in the Cancer Imaging Archive \citep{Clark2013_CTdata}. 
A set of candidate sequences representing diverse geometry functions was produced by randomly casting one hundred thousand rays through the CT geometry and tracing the distances each ray traversed through each distinct material and its density. In \figref{fig:megafigure}(a) this CT phantom is shown together with material sequences obtained from rays crossing at least \SI{30}{cm} of material (to exclude glancing trajectories), ordered in descending total mass thickness. 

From this set of sequences, 47 were randomly selected and used to perform proton beam simulations, each featuring a unique starting energy between \SI{70}{MeV} and \SI{250}{MeV}. The only condition constraining the selection was that the resulting material sequences should have sufficient mass thickness to stop each proton, as determined by the NIST PSTAR utility for protons stopping in lucite \citep{NIST_STAR_ref}, and with a small extra margin. 
The simulations consist of a proton beam incident on a cylindrical phantom, which in all cases has a radius of \SI{25}{cm} and a nominal maximum length of \SI{40}{cm}. This length is divided into $k_{max}$ sections, which were set to be 800 at most, each with a fixed width of \SI{0.5}{mm}. Following the proposed framework of subsection \ref{subsec:surrogate_design}, each section features lateral homogeneity with a different material composition and density identified by the selected series of functions $a_{k}$. The beam is in line with the axis of the cylinder and the spatial $z$-axis of the geometry, and is spawned \SI{70}{cm} upstream from the phantom's surface; the area outside of the phantom is filled with void/vacuum.  The beam has a radially symmetric Gaussian shape ($xy$) with a radial spread of standard variation $\sigma_{xy}=\SI{4.0}{mm}$ and no initial angular divergence.  
Moreover, the initial spread in energy is assumed to be Gaussian with energy-dependent width, pictured in absolute terms (FWHM) and relative terms in \figref{fig:megafigure}(b). This width was computed using the approach detailed in \citet{Tjelta2023}, but using updated fit parameters provided from calibration work at the proton therapy facility at Haukeland University Hospital in Bergen, Norway. 

From such simulations, ground truth phase space distributions $\psi_{k}^{p}$ and $\psi_{k}^{n}$ were collected through binning of transported protons and produced neutrons at each of the $k_{max}$ sections as described in subsection \ref{subsec:surrogate_design}. This was done by employing customized PHITS tallies, allowing versatile scoring of the particles and subsequent binning into suitable discretizations of the phase space domain, which we denote as $D^{p}$ and $D^{n}$ for protons and neutrons, respectively. Details of both discretizations are provided in table \ref{tab:discretization_details}.

\begin{table}[htbp]
  \centering
  \renewcommand{\arraystretch}{1.5}
  \begin{tabular}{cc|cc|cc|cc|cc}
    \hline
    \multirow{2}{*}{\textbf{Dimension}} &
    \multirow{2}{*}{\textbf{Unit}} &
      \multicolumn{2}{c|}{\textbf{Min.}} &
      \multicolumn{2}{c|}{\textbf{Max.}} &
      \multicolumn{2}{c|}{\textbf{Bins}} &
      \multicolumn{2}{c}{\textbf{Log-space}} \\
    \cline{3-10}
      & &
      $D^{p}$ & $D^{n}$ &
      $D^{p}$ & $D^{n}$ &
      $D^{p}$ & $D^{n}$ &
      $D^{p}$ & $D^{n}$ \\
    \hline
    \textbf{R}        & mm      & 0.18 & 0.0 &  95.9 &  60.0 &  30 &  30 &   yes   &   no   \\
    \textbf{E}        & MeV     & 0.0  & 0.0 & 250.0 & 250.0 & 250 & 125 &   no    &   no   \\
    \textbf{$\theta$} & \degree & 0.47 & 0.0 &  58.76& 180.0 &  30 &  30 &   yes   &   no   \\
    \hline
  \end{tabular}
  \renewcommand{\arraystretch}{1}
  \caption{Details of discretizations $D^{p}$ and $D^{n}$. For logarithmically spaced dimensions, the domain bounds were chosen as the $0.01\%$ and $99.9\%$ quantile values from a reduced dataset specifically collected for such purpose, with the first bin being extended to also include 0.}
  \label{tab:discretization_details}
\end{table}

As shown from the table, $D^{p}$ features logarithmic spacing along the $R$ and $\theta$ dimensions. These logarithmically spaced discretizations were employed in order to achieve higher resolution in regions of the domain where more particles are typically located, thereby mitigating the problems of having near-delta functions within collected observations and of resolving effects hidden beneath the discretization. However, the same was not applied along the $E$ dimension as it is spanned more evenly among different simulations and steps $k$, while a coarser energy binning coupled with linear spacing on $R$ and $\theta$ was used in $D^n$. These differences in the neutron discretization were taken because of the more spread-out shape of the $\psi_{k}^{n}$ distributions across the domain and the lower amount of overall counts, which would otherwise lead to an under-population of regions with finer resolution.

\begin{figure}[htbp]
	\centering
	\embedvideo{\includegraphics[width=1.00\textwidth, trim={0 0 0 0},clip]{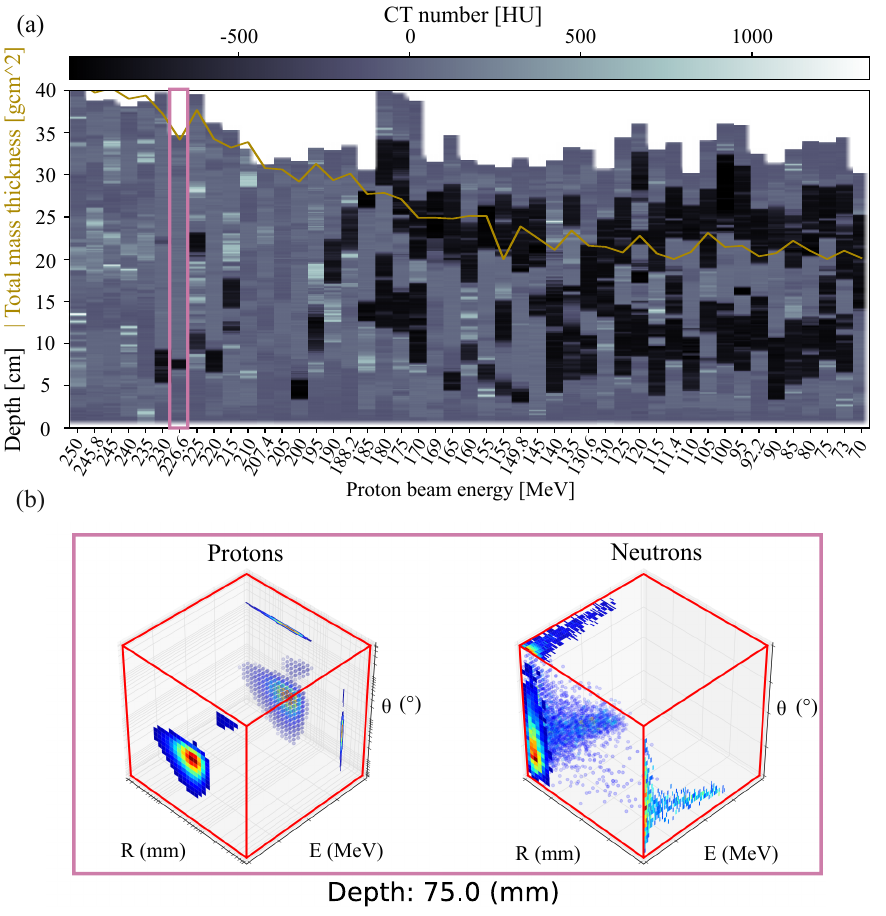}}{data_visu_transpose.mp4}
	\caption{ Visualization of the dataset: (a) the selected geometry function per depth shown for each simulated energy within the dataset; (b) examples of MC-binned densities $\psi^{p}_{k}$ and $\psi^{n}_{k}$ taken from the simulation highlighted in (a), showing only bins having a probability mass greater than $0.1\%$ and with projections of the distributions along each pair of dimensions plotted on the domain boundaries. The evolution over each step $k$ is also animated within this figure.
    }
	\label{fig:density_per_depth}

\end{figure}

The simulations thus performed were separated into a training and a test dataset on the basis of the starting energy. The training data consisted of 37 different energies, from \SI{70}{MeV} to \SI{250}{MeV} in \SI{5}{MeV} steps; the test data consisted of 10 different energies, spanning from \SI{73}{MeV} to \SI{245.8}{MeV} in \SI{19.2}{MeV} steps. Examples of the collected data are shown in \figref{fig:density_per_depth}, along with the selected geometry functions across the entire dataset.

The figure shows that collected samples of $\psi_{k}^{n}$ can present significant noise in regions associated with a low CT number. For this reason, each of the 47 simulations was performed twice, the first with one hundred million ($10^8$) proton histories simulated and another with one billion ($10^9$) proton histories simulated. This approach allowed the collection of approximated produced neutron densities $\psi_{k}^{n}$ with two different levels of statistical noise. We will refer to these two datasets as ES8 (energy spread, $10^8$ primaries) and ES9 (energy spread, $10^9$ primaries), thereby featuring a different amount of neutrons scored (although equally distributed), but the same proton phase space data since noise was deemed negligible in that case.

For further details on the enabled physics, nuclear data, tallying and scoring structure, the reader is referred to \ref{app:A}.

\subsection{AI methods}
\label{subsec:ai_methods}

\begin{figure}[htbp]
	\centering
	\includegraphics[width=1.00\textwidth, trim={0 0 0 0},clip]{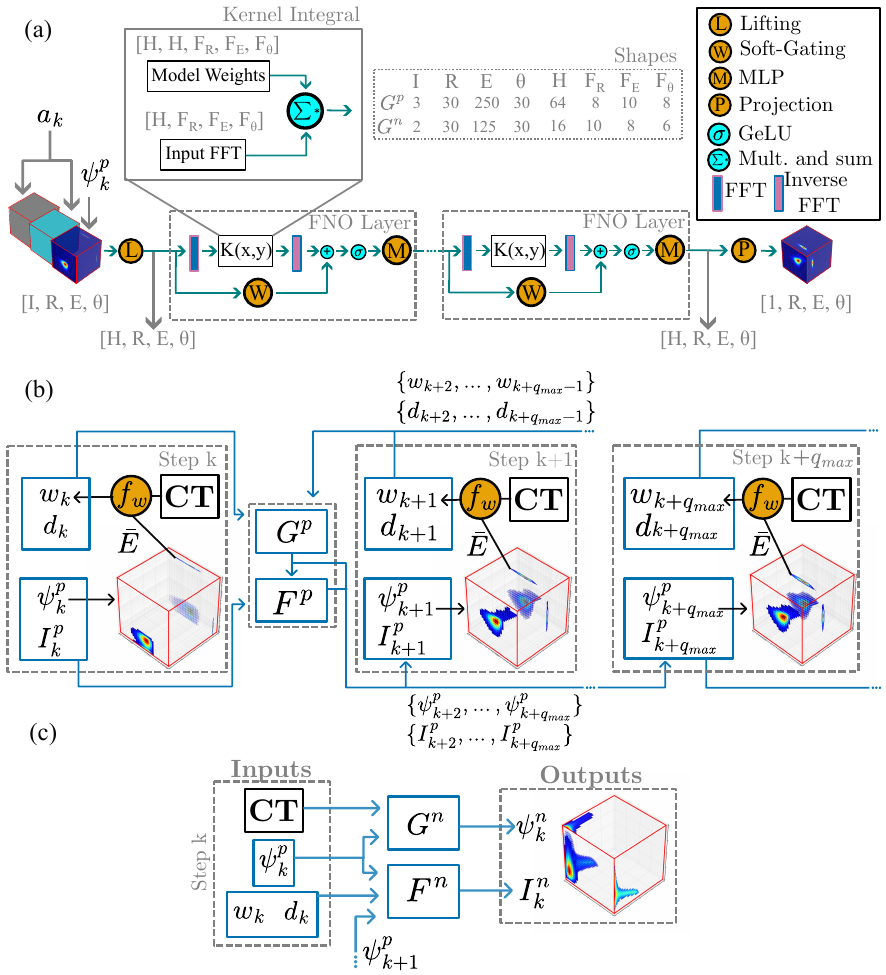}
	\caption{
    Details of the proposed surrogate: (a) the FNO architecture employed to implement $G^{p}$ and $G^{n}$ is shown, along with the shape of the data and network weights through its depth; (b) inter-operation between proton data and the multi-step surrogate components $G^{p}$ and $F^{p}$, where $f_{w}$ represents the mapping from the proton beam energy and CT number to the water equivalent thickness $w_k$; (c) inter-operation between the predicted proton phase space and the $G^{n}$ and $F^{n}$ components producing the neutron phase space prediction. 
    }
	\label{fig:surrogate_and_arch}

\end{figure}

The architecture employed for both $G^{p}$ and $G^{n}$ and their inter-operation with $F^{p}$ and $F^{n}$ are displayed in \figref{fig:surrogate_and_arch}. 

The FNO architecture depicted in \figref{fig:surrogate_and_arch}(a) was chosen due to its efficiency in handling high-dimensional discretizations, its robustness with respect to varying levels of noise, and its demonstrated performance in modeling time-evolving PDEs, which is achieved through a combination of local operations leveraging from a point-wise context together with global operations utilizing the entire input domain. In the case of FNOs, such global operations are implemented as point-wise multiplications between the FFT of the input functions and the model weights, which represent learned functions in Fourier space. Truncating the high-frequency components from such functions leads to an effective and cost-efficient convolutional layer capturing the full context from the input functions and limiting the effect of noise. The whole model consists of 4 such consecutive Fourier layers, with hyperparameters (i.e. the number of retained Fourier modes and the channel dimension) set as shown in the figure, and each complemented with soft-gating skip connections and additional channel-wise Multi-Layer Perceptrons (MLP) applied after activations. To control the channel dimension within hidden layers and at the output, we also employ lifting and projection operators in the form of $1\times1\times1$ convolutions. Following these specifications, the proposed $G^{p}$ and $G^{n}$ contain respectively around thirteen million and seven hundred thousand parameters. Moreover, we address the non-periodicity of the data and energy-dependence of the operators by applying zero-padding before each FFT, but we avoid exploring zero-shot super-resolution scenarios with our non-uniform discretizations due to the high cost of producing our data. 

As also anticipated in equations \eqref{eq:prot_op_zdiscr} and \eqref{eq:neutr_op_zdisc}, the input functions to both operators are the incident proton phase space $\psi_{k}^{p}$ and the geometry function $a_{k}$, which we provide to the models along the channel dimension by broadcasting it to the same size as the phase space discretization. 
In the case of $G^{n}$, we parametrize each $a_{k}$ simply as the CT number of the traversed material. However, we follow a different approach in the proton task, since the recursive task lasting up to 800 steps (i.e. the value of $k_{max}$) can lead to a considerable error accumulation in deeper regions of the phantom.
To address this problem, we instead parametrize each $a_k$ into two distinct functions $d_{k}, w_{k} \in\mathbb{R}$ respectively representing the absolute thickness and the energy-dependent water equivalent thickness \citep{zhang_calculation_2009} associated with step $k$, with the latter being computed from the mean energy of the incident protons, the material composition and density as defined from \figref{fig:megafigure}(b), and the energy-material stopping power tables from the PSTAR library \citep{NIST_STAR_ref}. We then use functions $w_{k\rightarrow k+q} = \sum_{i=k}^{k+q} w_{i}$ and $d_{k\rightarrow k+q} = \sum_{i=k}^{k+q} d_{i}$ as the supplementary inputs of $G^{p}$, and train it to perform its predictions over a variable number of steps up to a predefined $q_{max}$, allowing to divide the maximum amount of recursive operations to simulate the entire depth by such a factor. 

The implementation of functionals $F^{p}$ and $F^{n}$ instead utilizes much simpler models, although $F^{p}$ performs its prediction over a variable number of steps similarly to $G^{p}$. In both cases we implement them using gradient boosted tree regressors, and we set the input as the first two statistical moments of $\psi_{k}^{p}$ along $E$ and $\theta$, those of either $\psi_{k+q+1}^{p}$ or $\psi_{k+1}^{p}$ depending on the functional, and the material information in the form of absolute and water equivalent thickness traversed as scalar values. In both models we employed the same tree depth of 5, but set the number of trees per model to 512 and 256 for protons and neutrons respectively.

In \figref{fig:surrogate_and_arch}(b), the surrogate's operation over multiple steps concerning the proton task is shown, highlighting the inter-operation between $G^{p}$, $F^{p}$, and the proton phase space densities over depth. Here, each $w_{k}$ is computed using the CT number of the material to be traversed together with the average energy of protons at the previously predicted step. Starting from the predicted proton phase space, the CT number, and the pairs $w_{k}$ and $d_{k}$ at each individual step, $G^{n}$ and $F^{n}$ then compute the neutron density distributions and related intensity as shown in \figref{fig:surrogate_and_arch}(c). 

\subsection{Training details}
\label{subsec:training_details}

Both FNO architectures are implemented as explained in Subsection \ref{subsec:ai_methods} through the Python library \texttt{neuraloperator} \citep{kossaifi2024neural,kovachki_neural_2024}. Unlike $G^{p}$, the $G^{n}$ operator pre-processes the input data through a linear interpolation along $E$ to match the size of each input $\psi_{k}^{p}$ to the one used by discretization $D^{n}$, and mitigates noise by applying a smoothing Gaussian kernel to the ground truth maps $\psi_{k}^{n}$.

In both cases we used the Mean Absolute Error (MAE) as the training loss, and employed the Adam optimizer with a batch size of 4 and a starting learning rate of 0.001, decaying by a factor of 2 using a patience mechanism of 5 epochs. We trained for at most 200 epochs, and set an early stopping condition of 10 epochs. Model selection and both patience mechanisms were based on the MAE achieved on a validation set of 7 simulations, extracted from the training dataset based on the starting energy. 
In the case of $G_{p}$, input and target training pairs were extracted for all possible combinations of steps within a maximum distance of 10 steps, and patience mechanisms were set to track errors under recursive conditions. However, due to the considerably lower amounts of scored particles in regions near the proton range, the pairs were constructed only up to depths associated with an intiensity equal to $1\%$ of the primary one for $G^p$, and with an additional constraint of having scored at least 100 neutrons within the step for $G^{n}$.

Functionals $F^{p}$ and $F^{n}$ were trained separately using library \texttt{XGBoost} \citep{Chen_2016}: the former learned the relative reduction in the number of transported protons with respect to the incident ones, 
with a maximum reach of 25 steps; the latter would instead learn the absolute number of scored neutrons, normalized by the employed primary intensity. In both cases, we used the Mean Squared Error (MSE) as the training criterion, with a learning rate of 0.1. 

As mentioned in section \ref{subsec:data_collection}, two datasets with different primary intensities (named ES8 and ES9) were collected to assess the effect of statistical noise in the data caused by different amounts of scored particles. We denote the two surrogates trained on them as $M_{ES8}$ and $M_{ES9}$, differing solely in the operators $G^{n}$ since noise was deemed negligible in the proton distribution prediction and the two intensity prediction tasks. More details on the training and the resulting model weights can be found in the code and data repositories created for this paper \citep{Blangiardi_ai_2026, Blangiardi_Rodare}.

\subsection{Evaluation details}
\label{subsec:evaluation_metrics}
To evaluate the accuracy in the prediction of the phase space densities $\psi_{k}^{p}$ and $\psi_{k}^{n}$ and to guide model selection, for each ground truth-prediction pair we compute the total absolute error by summing the per-bin distances across the entire maps, dividing it by a factor of 2, and then averaging over predictions from different simulations and values of $k$. We refer to this metric as Probability Density Function-Mean Absolute Error (PDF-MAE), as it can be expressed in percentage of misplaced probability mass between the two distributions and lies in the range $[0, 100]\%$. 

In addition, we employ the Wasserstein distance between the predicted and ground truth distributions as our main metric, as it also captures the distance at which misplaced mass is scattered across the phase space domain. The intuitive interpretation of such distance corresponds to the minimum amount of ``physical work'' required to transport probability mass across the domain of the predicted density so that it is equal to the ground truth one. 
We compute it through the \texttt{SamplesLoss} method of the library \texttt{geomloss} \citep{feydy2019interpolating}, and express it in $\%_{frac} \times \mathbf{u}$, where $\mathbf{u}$ represents a distance in bins across either $D^{p}$ or $D^{n}$, meaning that the metric can attain a maximum value of $\approx 254$ and  $\approx 132$ in the proton transport and neutron production tasks respectively (i.e. $1.0$ times the length in bins of the discretization's longest diagonal).

To evaluate the regression task performed by the functionals, the predicted absolute intensities are compared with the ground truth using a Capped Mean Absolute Percentage Error (C-MAPE) with a maximum value of $300\%$. This allows to evaluate performances fairly in the presence of fluctuating ground truth values, while also mitigating extreme outliers associated with regions with intensities approaching 0. Additionally, we also estimate the accuracy of the depth-wise intensity distribution by computing the absolute difference in Range Landmark $\Delta RL$ as in \citet{setterdahl_enhancing_2025}, where the Range Landmark is defined as the distribution's first statistical moment.

Finally, we evaluate the inter-operation between phase space operators and intensity functionals by concatenating all predicted densities over $k$, multiplying them with the related intensity, and by computing the relative $L^{2}$ discrepancy with respect to the ground truth phase space as the ratio between the $L^{2}$ norm of the per-bin residuals and that of the ground truth densities, averaged across simulations. We complement this metric with a purely spatial evaluation through the gamma passing rate $\gamma_{pr}^{2mm,2\%}$ \citep{dhakal_symmetric_2014}, computed with a \SI{2}{\%}/\SI{2}{mm} tolerance and 10\% normalization threshold similarly to \citet{xiao_prompt_2024}. We compute this in 2D using the library \texttt{pymedphys} \citep{Biggs2022}, by integrating the aforementioned concatenated distributions over the $E$ and $\theta$ dimensions, and resampling the R dimension onto a linearly-spaced grid along lateral dimension $y$ with 80 bins, a resolution of \SI{0.5}{mm} and spanning the [-20, 20]\SI{}{mm} range. 

Surrogates have been evaluated under recursive operation by providing as input only the primary intensity, the phantom's material composition, and the incident proton density $\psi_{0}^{p}$. Accuracy metrics have been collected up to depths limited by the same intensity constraints as in the training procedure.

\section{Results}
\label{sec:results}

\subsection{Accuracy performance}

We treat surrogate $M_{ES8}$ as the main solution, and we evaluate its performance on both the $ES8$ and $ES9$ datasets. We also show the evaluation of $M_{ES9}$ on its respective dataset to quantify the improvement in performance achieved over $M_{ES8}$. A summary of the attained metrics discussed in Subsection \ref{subsec:evaluation_metrics} is shown in Table \ref{tab:results}, separated between the two datasets used for evaluation, the model performing the estimation, and the target phase space evaluated.

\begin{table}[htbp]
\centering
\renewcommand{\arraystretch}{1.5}
\begin{adjustbox}{max width=\textwidth}
\begin{tabular}{l c|*{3}{cc}|*{3}{cc}}
\hline
\multirow{2}{*}{\textbf{Model}} 
  & \multirow{2}{*}{\textbf{Target}}
  & \multicolumn{2}{c}{PDF-MAE} 
  & \multicolumn{2}{c}{Wass. Dist.} 
  & \multicolumn{2}{c|}{C-MAPE} 
  & \multicolumn{2}{c}{$L^{2}$ Error}
  & \multicolumn{2}{c}{$\gamma_{pr}^{2mm,2\%}$} 
  & \multicolumn{2}{c}{$\Delta RL$} \\
\cline{3-14}
  & 
  & \multicolumn{2}{c}{[$\%$]}
  & \multicolumn{2}{c}{[$\%_{frac.} \times u$]}
  & \multicolumn{2}{c|}{[$\%$]} 
  & \multicolumn{2}{c}{[frac]} 
  & \multicolumn{2}{c}{[$\%$]}
  & \multicolumn{2}{c}{[$mm$]}\\
\hline
\multicolumn{2}{c|}{Evaluation on $ES8$} 
  & Mean & $Q_{90}$
  & Mean & $Q_{90}$
  & Mean & $Q_{90}$
  & Mean & SD
  & Mean & SD
  & Mean & SD \\
\hline
$M_{ES8}$ & $\psi^{p}$   & 4.67 & 9.73 & 0.494 & 0.458 & 2.23 & 0.40 & 0.067 & 0.037 & 99.95 & 0.10 & 0.238 & 0.161 \\
$M_{ES8}$ & $\psi^{n}$   & 21.27 & 33.90  & 0.710 & 0.842 & 9.46 & 11.91 & 0.269 & 0.029 & 99.38 & 0.82 & 0.718 &  0.627 \\
\hline
\multicolumn{2}{c|}{Evaluation on $ES9$} 
  & Mean & $Q_{90}$
  & Mean & $Q_{90}$
  & Mean & $Q_{90}$
  & Mean & SD
  & Mean & SD
  & Mean & SD \\
\hline
$M_{ES8}$ & $\psi^{n}$   & 10.60 & 18.33 & 0.709 & 0.695 & 11.51 & 13.04 & 0.137 & 0.034 & 99.40 & 0.80 & 0.871 & 0.610\\
$M_{ES9}$ & $\psi^{n}$   & 8.73 & 13.33 & 0.507 & 0.484 &  - & - & 0.121 & 0.030 & 99.43 & 0.77 & - & - \\
\hline
\end{tabular}
\end{adjustbox}
\renewcommand{\arraystretch}{1}
\caption{
Collected evaluation metrics for each model and dataset, separated by target phase space density. For the PDF-MAE, the Wasserstein Distance and C-MAPE, metrics are aggregated over all steps of all simulations; however, for the $L^{2}$ error, $\gamma_{pr}^{2mm,2\%}$ and for $\Delta RL$ the aggregation is computed solely across simulations, as they are not computed for each step. We therefore show the mean value and the 90\% quantile in the former cases, and the mean and standard deviation (SD) in the latter. As $M_{ES8}$ and $M_{ES9}$ share the same $G^{p}$, $F^{n}$ and $F^{p}$, their respective metrics are only reported once.
}
\label{tab:results}
\end{table}

\begin{figure}[htbp]
	\centering
	\includegraphics[width=1.0\textwidth, trim={0 0 0 0},clip]{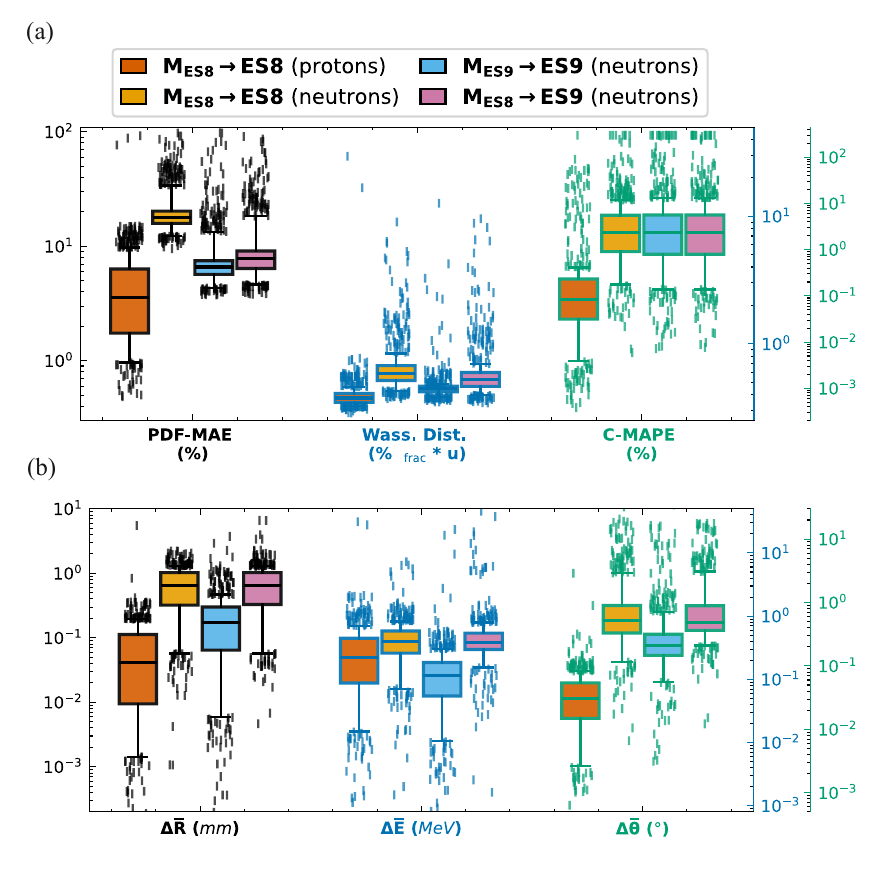}
    \caption{
    Box plots showcasing errors of the trained surrogates at each step of all evaluated simulations: (a) error distributions for the PDF-MAE, Wasserstein distance and C-MAPE metrics for the predicted densities; (b) mean absolute distance between the mean values of the predicted distributions along $R$, $E$ and $\theta$. Shown for both protons and neutrons for every evaluated combination of models and datasets. For each box, horizontal lines indicate the $25^{th}$, $50^{th}$ and $75^{th}$ quantile values, while whiskers indicate the $5^{th}$ and $90^{th}$ ones. Data points outside this range are undersampled randomly and shown as vertical markers.
    }
	\label{fig:box_plots_errs}

\end{figure}

Overall, the surrogate is able to generate considerably accurate distributions for both protons and neutrons. This can be seen from the Wasserstein distance being below $1.0\%_{frac} \times \textbf{u}$ in all cases, indicating that the work required to make the predicted distribution match the ground truth is equivalent to that of displacing it by less than one bin. 
This leads to spatial distributions with very high $\gamma_{pr}^{2mm,2\%}$, exceeding 99\% for both proton and neutron densities.
Notably however, the Wasserstein distance, the PDF-MAE, and the $L^{2}$ relative error are higher for the neutron production task than for the proton transport. The performance degradation is sharper for the latter two metrics since they both evaluate per-bin distances, which makes them particularly sensitive to statistical noise within the evaluation data. The effect of such noise on them can also be observed in the better performance achieved by $M_{ES8}$ on the neutron task when evaluated on the less noisy $ES9$ dataset. An opposite trend is instead observed for the C-MAPE and $\Delta RL$ despite being produced by the same $F^{n}$ component, which is due to $ES9$ being also evaluated in low-intensity regions slightly deeper in the phantom.  
Moreover, it is also evident that $G^{n}$ can achieve better performance when trained on $ES9$, as $M_{ES9}$ consistently outperforms $M_{ES8}$ in the related neutron metrics.
In \figref{fig:box_plots_errs}(a), we visualize how the step-wise errors are distributed among all evaluated data points, while \figref{fig:box_plots_errs}(b) shows the related error along the computed mean of each dimension in the distribution domains. 

\FloatBarrier

From table \ref{tab:results} and \figref{fig:box_plots_errs}(a), it is possible to see that the Wasserstein distance and the C-MAPE present mean values that are either very close or above the $Q_{90}$ value, which is caused by extreme outliers that are several orders of magnitude above the majority of the distribution. From \figref{fig:box_plots_errs}(b), it is instead possible to see that the predicted densities maintain fairly accurate physical properties, as most of the predicted densities fall within an error in their first statistical moment below \SI{1}{mm} along $R$, \SI{1}{MeV} along $E$ and \SI{1}{\degree} along $\theta$, which is mostly within the resolution of the data coherently with the computed Wasserstein distance.

\begin{figure}[htbp]
	\centering
	\includegraphics[width=1.0\textwidth, trim={0 0 0 0},clip]{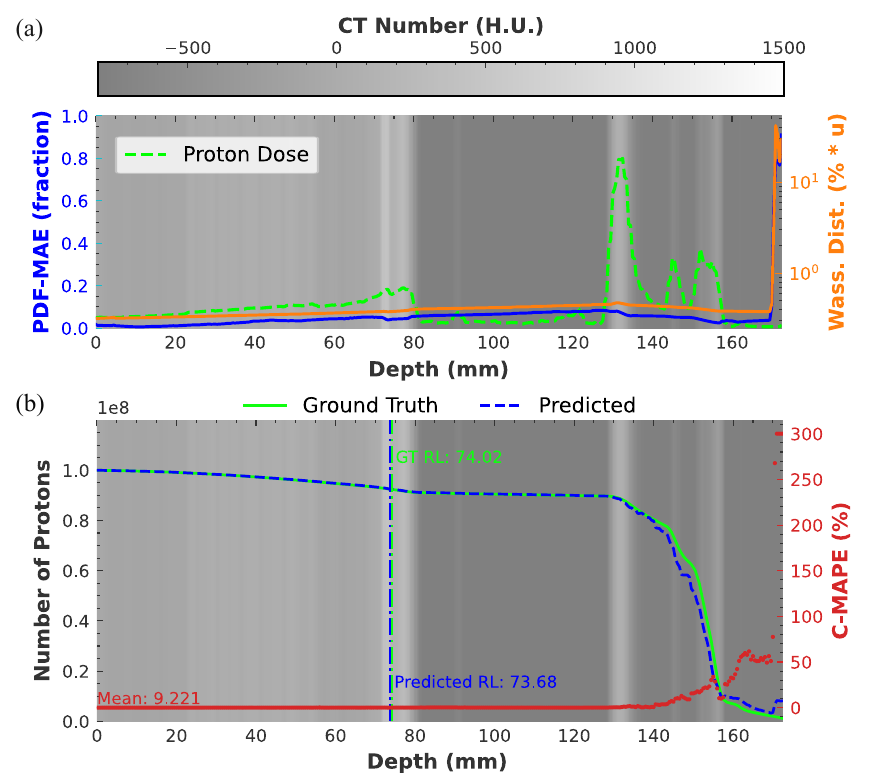}
	\caption{
    Error profile attained in the predicted proton phase space as a function of depth, shown for the simulation having the worst performance in terms of Wasserstein distance: (a) the PDF-MAE, Wasserstein distance and the energy deposited by the beam are displayed; in (b), predicted and ground truth intensity at each depth are compared, and the C-MAPE is displayed together with the range landmarks of the two distributions.
    }
	\label{fig:main_profile_protons}

\end{figure}

\begin{figure}[htbp]
	\centering
	\includegraphics[width=1.0\textwidth, trim={0 0 0 0},clip]{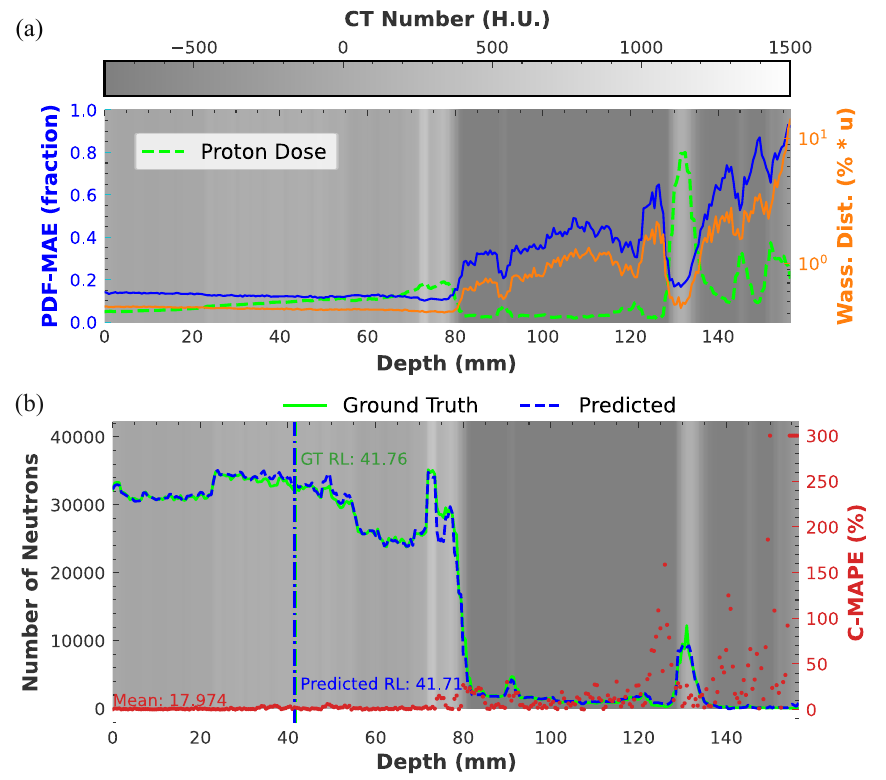}
	\caption{
    Error profiles attained in the predicted neutron phase space within the simulation with the highest Wasserstein distance. Follows the same structure as \figref{fig:main_profile_protons}.
    }
	\label{fig:main_profile_neutrons}

\end{figure}

\begin{figure}[htbp]
	\centering
    \embedvideo{\includegraphics[width=1.00\textwidth, trim={0 0 0 0},clip]{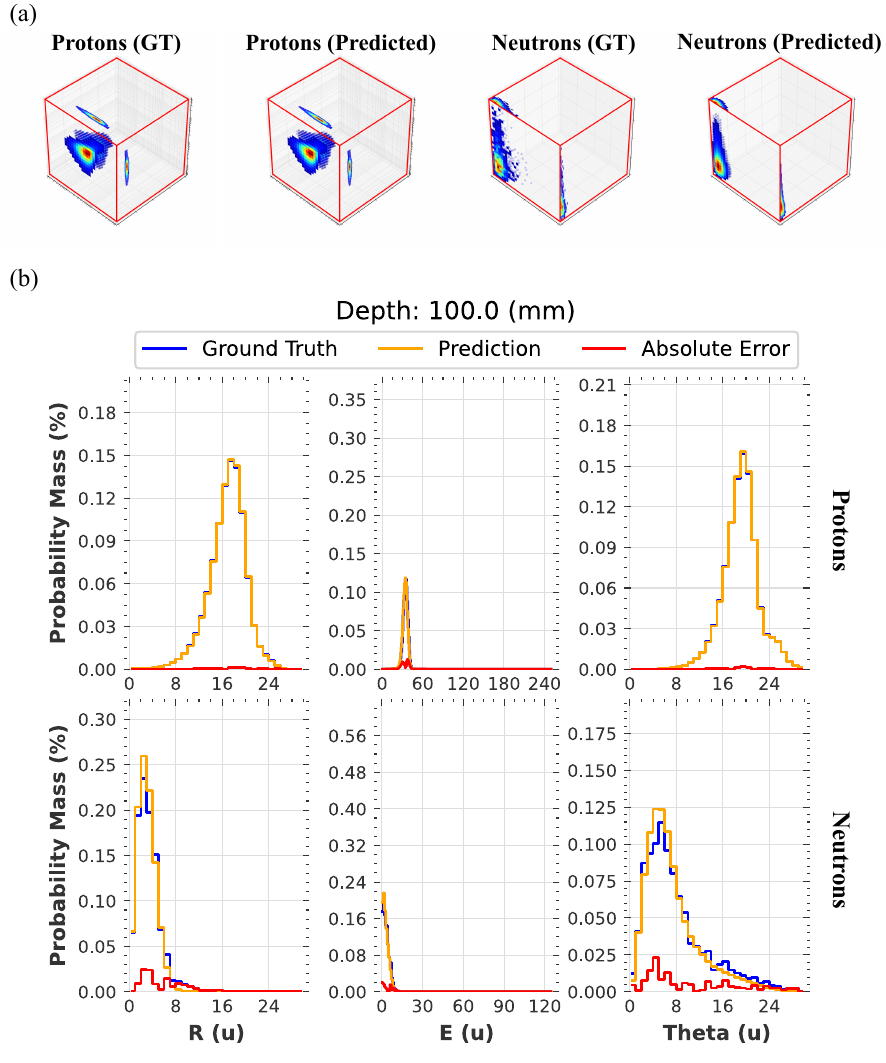}}{depth_inspection.mp4}
    \caption{Comparisons between predicted and ground truth $\psi_{k}^{p}$ and $\psi_{k}^{n}$, from the same simulation with a starting proton energy of \SI{111.4}{MeV} displayed in \figref{fig:main_profile_protons} and \figref{fig:main_profile_neutrons}: (a) full 3D ground truth and predicted distributions are shown for both neutrons and protons, following the same visualization as \figref{fig:density_per_depth}; (b) 1-dimensional projections along each dimension of both ground truth and predictions, along with the absolute error in fractional percentage. This figure also animates the full evolution in depth of such visualizations.}
	\label{fig:depth_inspection}

\end{figure}

We now provide a visual indication of both of these effects separately for proton and neutron densities in \figref{fig:main_profile_protons}, \figref{fig:main_profile_neutrons}, and \figref{fig:depth_inspection}. The figures show predicted distributions and errors attained by $M_{ES8}$ within the simulation with a starting energy of \SI{111.4}{MeV}, which featured the highest average Wasserstein distance for both types of densities when evaluated on $ES8$.

As shown by the deposited energy, this simulation consists of a proton beam crossing and stopping in a region primarily composed of low-density lung tissue, with the dose deposition curve not showing a clear Bragg peak due to the significant heterogeneity of the traversed tissues. Although this scenario is particularly challenging, \figref{fig:main_profile_protons} shows how the Wasserstein distance in the proton task attains values close to the global mean, except for a few steps at the end of the simulation that are two orders of magnitude higher. In \figref{fig:depth_inspection} it is possible to see that the overall discrepancies are fairly minimal, while the aforementioned high errors are due to $G_{p}$ not being able to map the proton phase space to a meaningful next state beyond the range, which is likely due to the lack of training data points in those regions of the phase space domain. However, it should be noted that those regions are also associated with few protons, few secondary particles and low delivered dose, meaning that the aforementioned errors have limited relevance for practical applications since they can be easily detected and managed. 

\begin{figure}[htbp]
	\centering
	\embedvideo{\includegraphics[width=1.00\textwidth, trim={0 0 0 0},clip]{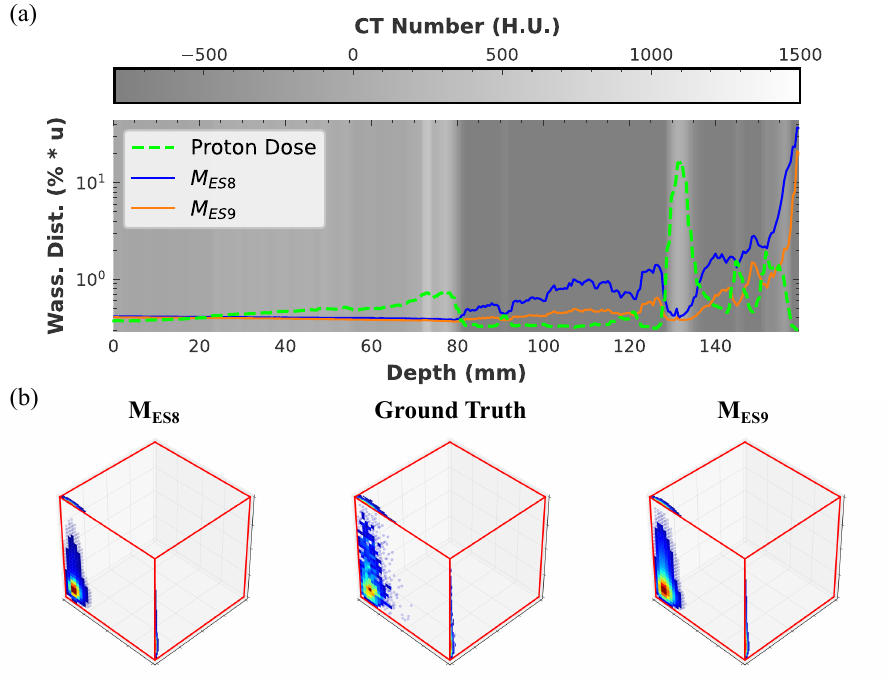}}{1e9_comparison.mp4}
    \caption{
    Evaluation of the predicted densities $\psi_{k}^{n}$ on the \SI{111.4}{MeV} simulation from the $M_{ES9}$ dataset: (a) shows the Wasserstein distance over depth attained by both $M_{ES8}$ and $M_{ES9}$; (b) visualizes the full distributions predicted by the two models together with the corresponding ground truth.
    Animations within this figure also show the predicted and ground truth density functions at each step $k$.
    }
	\label{fig:1e9_comparison}
\end{figure}

Differently from the proton task, the error profiles in the neutron density prediction depicted in \figref{fig:main_profile_neutrons}(a) show an inverse correlation between the errors and the CT number of the traversed material, resulting in error fluctuations of more than one order of magnitude. For the C-MAPE in \figref{fig:main_profile_neutrons}(b), this is mostly due to the ground truth number of particles reaching very low intensities, leading to mis-predictions being more severely punished. In the case of the phase space metrics, however, part of the reason can be found in the higher statistical noise in the ground truth phase space, which is more evident in the neutron densities displayed in \figref{fig:depth_inspection}. To support such consideration, we compare the performance of both $M_{ES8}$ and $M_{ES9}$ when evaluated on $ES9$ in \figref{fig:1e9_comparison}.

When compared to the evaluation on $ES8$ in \figref{fig:main_profile_neutrons}(a), the Wasserstein distance achieved by $M_{ES8}$ in $ES9$ is relatively lower in the lung region despite the predicted densities being the same, confirming that noise in the evaluation data is affecting the reported metrics. However, it is clear that using higher-intensity data also during training leads to improved performance, as within this specific simulation $M_{ES9}$ achieves an average Wasserstein distance of only 0.66 $\%_{frac.} \times u$ as opposed to the 1.19 $\%_{frac.} \times u$ achieved by $M_{ES8}$. As shown in \figref{fig:1e9_comparison}(b), such differences in error manifest in smaller, more localized predictions in the case of $M_{ES8}$, leading to under-predictions in the tails of the distributions. 
Since few particles are normally scored in low-density materials within $ES8$, this is likely due to the FFT systematically filtering out as noise the even sparser production scored in the outer regions of the beam. 
However, precisely because they happen together with mostly negligible production rates, these effects do not reflect as negatively in metrics applying normalization on the entire depth-wise distributions, such as the $\gamma_{pr}^{2mm,2\%}$ and the $L^{2}$ discrepancy, as indicated by table \ref{tab:results} where their degradation between $M_{ES8}$ and $M_{ES9}$ is not as significant.

\subsection{Time performance}

Simulations were conducted on several High Performance Computing (HPC) cluster nodes, each featuring an Intel Xeon Gold 6148 CPU (20 cores, 40 threads, \SI{2.40}{GHz} base frequency). PHITS version 3.341 compiled and executed with OpenMP shared-memory parallelization was employed with 10 threads utilized per job. \ref{app:A} further details the computational cost of these simulations, which ultimately resulted in simulation times of 14.7 and 27.0 CPU years for the entirety of $ES8$ and $ES9$, respectively. 

The time performance of the AI methods was measured on a single node with an NVIDIA A100 GPU and a \SI{3.8}{GHz} AMD EPYC 9354 32-Core Processor. The surrogate was set to predict proton and neutron phase space distributions for the entire 800 steps of the phantom, regardless of the specific beam range, and sequentially between the proton and neutron tasks. On average, performing each simulation required a compute time of \SI{23.17}{s}, of which \SI{17.82}{s} were spent on GPU-related computations, and divided into \SI{11.15}{s} and \SI{6.67}{s} respectively for the proton and neutron tasks. This leads to a considerable speed-up of several orders of magnitude with respect to the employed MC code.

\section{Discussion}

\subsection{Significance}
Within this work, we proposed a fast proton transport and neutron production surrogate model for proton therapy using FNOs and gradient boosted trees. We have considered depth as a pseudo-time dimension, divided the remaining phase space into a subset of relevant dimensions, and 
learned its evolution over depth in a data-driven way for both protons and neutrons.
Unlike other deep learning task-specific methods, our surrogate explicitly predicts proton distributions auto-regressively using only a local context, and with no information on the length of the sequence. This allowed the model to learn the task with a relatively low amount of training data and retain interpretability among predicted steps.  
We targeted angular- and energy- resolved proton distributions because of their correlations with those of produced neutrons, which we have subsequently computed using a similar model to the one used for protons. 

The fine resolutions employed, both in the models' input and the phantom's depth, did not compromise the accuracy achieved, as the chosen network can handle the high dimensionality in a lower-dimensional Fourier space while mapping proton distributions across variable depths. 
Our study on laterally heterogeneous phantoms confirms this as the predicted distributions achieved average Wasserstein distances with respect to MC data of only $0.494$ and $0.710$ $\%_{frac} \times \mathbf{u}$ in the proton and neutron task respectively, indicating an overall misplacement within a single bin in optimal transport sense, and leading to an average spatial $\gamma_{pr}^{2mm,2\%}$ above $99.3\%$ in all cases.
As our data is obtained from binning of MC particles, we have observed that training the neutron model on a dataset produced with $10^{9}$ protons has decreased the average Wasserstein distance by $\sim30\%$, and the relative $L^2$ discrepancy by $\sim12\%$ with respect to using a primary intensity of $10^{8}$. Within our experiments, simulating such higher primary intensities led to a significant increase in computational costs for data generation, meaning that a suitable trade-off should be identified depending on the end application of the surrogate.

Reported metrics have been collected while testing the robustness of the model to different material sequences and initial proton energy and spread. With sufficient data, we believe the model to be able to robustly generalize over material sequences extracted from any anatomical site or patient without need for specific retraining, while also having the potential to handle changes to the beam's initial angular and spatial distributions through minimal finetuning.
Our experiments indicate that our solution requires $\sim$\SI{23}{s} to perform its predictions over \SI{40}{cm} at steps of \SI{0.5}{mm}, of which $\sim$\SI{18}{s} are required for the sequential model forwarding of the proton and neutron phase space models. Since this represents a considerable speedup with respect to the employed MC code, we believe our approach to be relevant for prototyping and deployment of multi-particle range verification pipelines, estimation of neutron dose, and prediction of further emissions from neutron-induced reactions, while also being easily adaptable for other types of secondary emissions within proton therapy.

\subsection{Comparison with other works}
Due to the assumptions made in this work and the largely unexplored modeling of neutrons in IMPT for range verification, a direct comparison of the proposed solution to previous research is not straightforward.

Most works accelerating the proton transport evaluate their solutions on the resulting spatial dose maps, which is not the objective of this work. In practice, the multi-step design chosen for the proposed proton transport surrogate is unsuited for such computations, as small differential changes between subsequent steps have to be resolved by the model without any direct context of the first step.
Therefore, dose-specific deep learning solutions such as \citet{pastor-serrano_millisecond_2022}, achieveing $\gamma_{pr}^{2mm,2\%}$ above $99.73\%$ even in highly challenging lung irradiation plans, remain the most suitable solutions for fast dose predictions despite our similarly accurate $\gamma_{pr}^{2mm,2\%}$ in the protons' spatial distributions. However, such solutions do not provide detailed angular and energy distributions of the protons, which we have set as one of our objectives. 

Therefore, we find works such as \citet{stammer_deterministic_2025} to be more closely aligned with ours. They develop a deterministic solver for proton transport using a dynamic low-rank approximation, modeling the LBTE under CSDA while targeting fine angular resolutions and reduced computational and memory costs. 
The reported solvers were benchmarked against the TOPAS MC code in proton transport simulations within a water tank featuring an optional lateral air cavity, achieving an $L^{2}$ relative error in the spatial dose distributions ranging between $0.039$ and $0.067$, depending on the employed angular resolution and rank approximation.
The runtime of the proposed solvers was stated to be $1\%$ of the full-rank solver described in \citet{kusch_robust_2023}, which required \SI{5408}{s} on a grid of $40000$ spatial bins and a rougher angular resolution than those employed in the low-rank approximation. Although our operator $G^{p}$ doesn't account for lateral heterogeneity, our reported $L^{2}$ norm of $0.067$ over the entire domain considered is comparatively in line with the described solution, while our proton runtime of \SI{11.15}{s} is shorter.

When it comes to neutron-specific works within proton therapy, the primary interest in their energy, spatial and angular distributions is due to their high radiobiological effectiveness \citep{howell_secondary_2014, anferov_analytic_2010, marafini_mondo_2017}. The work of \citet{shrestha_stray_2022} presents an analytical model for predicting neutron dose caused by production in the treatment nozzle during passively scattered proton therapy. The nozzle was modeled as the only neutron source, and the dose was computed using the spectral and angular fluence of the transported neutrons, which was estimated at every point by taking into account attenuation, divergence and scattering within the phantom. The model was benchmarked against MC simulations of an homogeneous water tank, and the quality of the fluence prediction was evaluated in terms of the distance between the means of the energy distributions, achieving an average relative distance between 16\% and 23\% depending on the beam energy. 

Instead, the objective of \citet{schneider_neutrons_2017} is the neutron dose estimation in the context of IMPT, which is done by utilizing range information of each pencil beam coming from a proton beam algorithm. Here, each point along the beam axis was treated as a neutron source, and a parametric formula was fitted with data from MC simulations to compute their contributions to the dose throughout the phantom. 
Quantitative evaluation of the parametrization on the fitted settings found an agreement in the neutron equivalent dose of $20\%$ in root mean squared error. In comparison to these works, our model is computationally more expensive and presents more constraints in resolution. However, it can handle heterogeneity and neutron production along the depth, unlike the model in \citet{shrestha_stray_2022}, and retains directional and energy components of the neutrons, which are collapsed into the spatial dose in \citet{schneider_neutrons_2017}, while it approaches MC accuracy.

Finally, we compare our solution with two other works that perform a similar emission prediction task, but target positron-emitters and prompt gamma rays. The first work \citep{pinto_filtering_2020} computes the number of emitted secondary particles by designing a filter function to be convolved with proton distribution parameters extracted from proton beam algorithms. In the case of prompt gamma rays, this was complemented with a look-up table approach to obtain emission spectra as well.
Benchmarks with MC on realistic datasets reveal normalized relative MAE on longitudinal profiles of up to 5.3\% in a lower abdomen treatment plan and estimated averaged shifts between \SI{-1.3}{mm} and \SI{0.9}{mm}. Secondly, \citet{xiao_prompt_2024} similarly aimed at predicting spatial emissions of prompt gamma rays using a long short-term memory deep learning architecture. Here, the network implicitly models the proton beam state through its memory mechanisms, and produces the prompt gamma emission maps throughout the phantom in sequential steps along the depth. The best model utilized MC-accurate dose maps as an additional input to the material composition, achieving an average $\gamma_{pr}^{2mm,2\%}$ of $98.5\%$ and an average shift of only \SI{0.15}{mm} when tested under variable proton beam energies in their prostate cancer CT dataset. In comparison to both of these solutions, our surrogate assuming lateral homogeneity achieves a C-MAPE of $9.46\%$ (which we have observed to be significantly affected by air cavities) an average $\Delta RL$ of \SI{0.871}{mm}, and an average $\gamma_{pr}^{2mm,2\%}$ of $99.38\%$, indicating that our solution is comparable in accuracy to those described. The key difference in our task of neutron prediction lies in the need to predict the angular emissions of the particles in addition to spatial and energy distributions, which have a correlation with both the energy and divergence of the proton beam. As we set ourselves to achieve MC-level accuracy, we hence estimate the neutron prediction through explicit auto-regression of the proton transport rather than by relying on proton beam algorithms or by directly learning the evolution of entire sequences. 
However, it should be noted that the higher dimensionality of our task due to the explicit angular and energy modeling (leading to each simulation being larger in number of bins by $\sim$2-3 orders of magnitude, comparatively to \citet{xiao_prompt_2024}) makes our solution slower than other task-specific deep learning methods, with our model requiring $\sim$\SI{18}{s} per beam spot as opposed to the $\sim$\SI{0.1}{s} reported by other solutions.

\subsection{Limitations and Future Work}
The main limitation of our current model lies in its assumption of lateral homogeneity of the phantom, which allowed us to remove one spatial dimension from the model. Although the current design can accommodate heterogeneity of the CT geometry along $R$, a study on its feasibility and further adjustments to include full 3D heterogeneity will be performed. 

Another limitation lies in the limited size of our dataset. Although we have specifically chosen highly heterogeneous scenarios to evaluate our surrogate, have designed a spatially local model with high generalization capabilities, and have conducted most of our evaluation at a local level, our evaluation dataset ultimately contains 10 proton beam simulations, all synthetically generated from a single thoracic CT scan. Therefore, a thorough evaluation on realistic phantoms employing lateral heterogeneity and from different anatomical sites will be performed in future works. An effective way to reduce costs in data production would be to avoid simulating high primary intensities, which were necessary in this work to obtain neutron data with sufficient statistical quality. Possible strategies allowing that include the use of variance reduction techniques, or a semi-analytical definition of the neutron operator following the look-up table approach of \citet{pinto_filtering_2020}.

Dataset limitations also led us to not explore zero-shot super-resolution properties of the employed FNOs, which normally allow to perform inference at different discretizations than those used during training. We therefore opted for a fairly simple form of the FNO, but more advanced formulations of such architecture can in fact be explored, such as variants addressing more directly the translational asymmetries in our task \citep{gao2025dynamic}, or other extensions effectively learning the solution operator in a discretization-convergent way also in our logarithmically spaced data \citep{li_fourier_2024}.

Finally, further improvements can be brought to the surrogate to decrease its inference time. Within this work, we have chosen the energy-dependent water equivalent thickness to represent material compositions within the proton model, meaning that 
each step in the phantom has to be simulated sequentially in order to make the mean energy available for the next step's computation.
However, simpler parametrizations of the geometry function could be explored to avoid this dependency, which would enable parallel computation of several steps at a time. 

\section{Conclusions}
As part of a novel range and dose verification concept in proton therapy making use of multiple secondary radiation types, we have developed a fast surrogate model to predict the phase space distributions of transported protons and produced neutrons, aiding in the pre-computation of the system's response. Our results on laterally homogeneous phantoms indicate that the model can achieve comparable accuracy to MC, with an average relative $L^2$ discrepancy of $0.067$ and $0.121$ for protons and neutrons respectively, while operating at a fraction of the computational time. We believe that our approach has the potential to enable clinical implementation of neutron-based range verification pipelines as well as other tasks requiring angular, energy and spatial distributions of neutrons produced within proton therapy.

\section*{Appendix A}
\label{app:A}

\begin{figure}[htbp]
	\centering
	\includegraphics[width=1.00\textwidth, trim={0 0 0 0},clip]{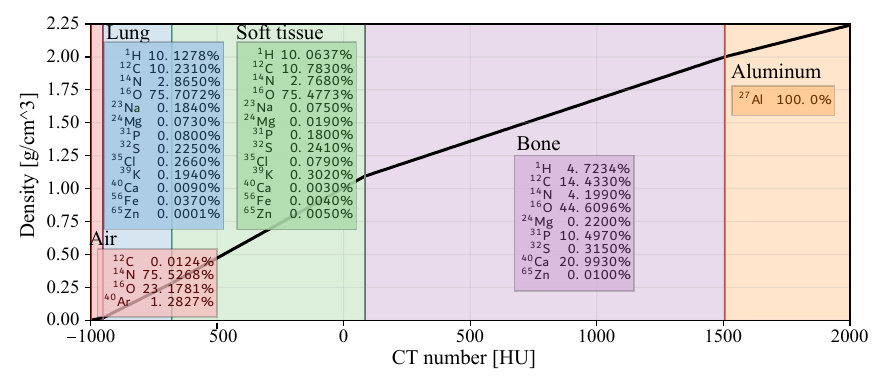}
	\caption{ Conversion between CT number in HU and material composition and density within PHITS. 
    }
    \label{fig:megafigure_material}
\end{figure}

To extract and simulate the sequences of selected material profiles within PHITS, the RT-PHITS utility \citep{RTPHITS_ref1of2,RTPHITS_ref2of2} was employed to convert the chosen DICOM file into a three-dimensional PHITS input geometry, and the whole range of attained CT values was converted from Hounsfield Units (HU) to individual material densities each corresponding to one of five material compositions \citep{Onizuka2016}, as illustrated in \figref{fig:megafigure_material}. 

The extraction was achieved with a custom-made PHITS tally, using a ``user-defined tally'' Fortran subroutine and recompiling PHITS. A cylindrical shell source (of radius \SI{25}{cm} and height \SI{40}{cm}) was placed around the CT phantom geometry (axially aligned with the CT vertical axis) and set to spawn neutrinos evenly across the surface and with random direction. Neutrinos were used to trace these rays since their interaction physics are disabled by default, meaning they pass through the geometry in a straight line without suffering any collisions in the problem geometry.  One hundred thousand neutrino rays were traced in this fashion. 

For the actual proton beam simulations on the extracted profile, settings enabled included Coulomb diffusion / angle straggling physics, energy straggling physics, and ``event generator mode'' default physics and models for more accurate modeling of event-by-event information \citep{emode_ref1,emode_ref2,emode_ref3}. The JENDL-5 nuclear data \citep{JENDL5_ref} were employed where available for relevant particles, along with the the INCL nuclear reaction model \citep{INCL46_ref} and the KUROTAMA model \citep{KUROTAMA_ref}.

\begin{figure}[H]
	\centering
	\includegraphics[width=1.0\textwidth, trim={0 0cm 0 0cm},clip]{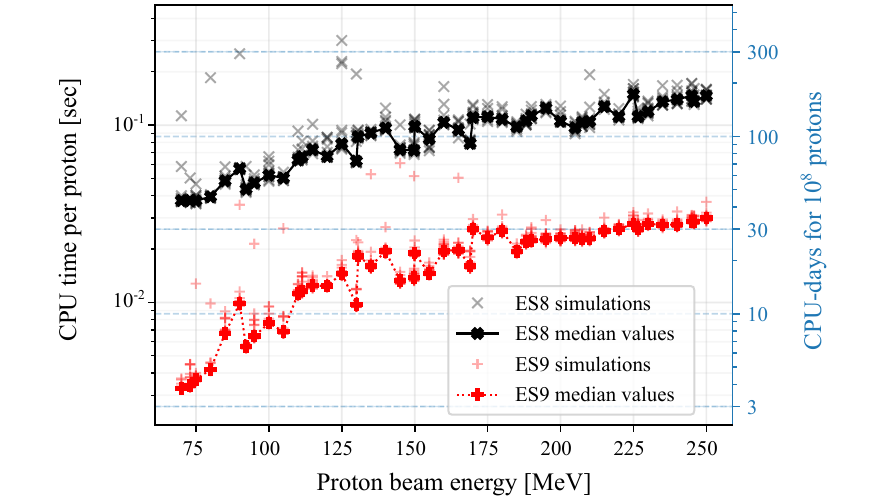}
	\caption{Summary of per-history-normalized CPU time required for each PHITS simulation, with ES8 simulations having all tallies (including the very large proton tally) enabled and the ES9 simulations only having neutron and gamma-ray production ``dump'' tallies enabled.  Each ES8/ES9 simulation was actually ran as 10/20 split simulations of $10^7$/$5\times10^7$ primary protons each, ran sequentially (picking up where the previous left off); the fainter symbols are for these 10/20 simulations conducted at each energy with the connected bolder symbols showing the medians of these times at each energy.}
	\label{phits-cpu-time}
\end{figure}

The proton phase space densities at each step $k$ were directly binned within PHITS using a customized tally implementing $D^{p}$ as specified in table \ref{tab:discretization_details}. As the resulting tally was quite large, enabling it significantly affected the runtime of the performed simulation, but was deemed as the more sensible solution in terms of storage size as list-mode scoring of each particle using the PHITS ``dump'' format was deemed to be intractable (in the order of tens of GB for simulations of only $10^{6}$ protons). As such, the data collection between $ES8$ and $ES9$ differed in the enabling of such large tally, but in both cases the ``dump'' functionality was used to store each generated neutron with an energy above \SI{10}{keV} in terms of full spatial, angular, energy and timing information, so that it could be used for binning in post-processing. Prompt gamma rays were also scored during simulations, whose list-mode data is available in \citet{Ratliff_Rodare}. The runtimes for the production of $ES8$ and $ES9$, resulting in a total of 14.18 and 27.0 CPU-years, is detailed in figure \ref{phits-cpu-time}

The PHITS Tools Python utility \citep{Ratliff2025_PHITSTools} was used to automatically parse all PHITS outputs---standard tally histogram files and binary tally ``dump'' files---and export them into compressed Pandas DataFrame objects utilized in AI training.

\section*{Acknowledgements}
The authors thank Dr. Stefan E. Müller at HZDR for his assistance with installation and troubleshooting of the PHITS environment on the HZDR computing resources used in this work, and we would like to thank our NOVO collaboration partner institution HZDR for providing the computational resources used for the PHITS simulations in this work.

During the preparation of this work, the authors used Grammarly, ChatGPT and related deep research tools in order to assist exploratory work and improve the readability of the manuscript. After using these tools, the authors reviewed and edited the content as needed and take full responsibility for the content of the published article.

\section*{Funding}
The NOVO project has received funding from the European Innovation Council (EIC) under grant agreement No. 101130979. The EIC receives support from the European Union’s Horizon Europe research and innovation program. Partners from The University of Manchester have received funding from UK Research and Innovation under grant agreement No. 10102118. 

Views and opinions expressed are, however, those of the author(s) only and do not necessarily reflect those of the European Union or the European Innovation Council and SMEs Executive Agency (EISMEA). Neither the European Union nor the granting authority can be held responsible for them.

\section*{Data and Code}
For this work's training and testing data, the raw files produced through PHITS \citep{PHITS333_ref} and initially processed by PHITS Tools \citep{Ratliff2025_PHITSTools} can be found at \citet{Ratliff_Rodare}, while the processed data used by the AI for each step $k$ and simulation can be found in \citet{Blangiardi_Rodare}. The repository implementing all relevant model-development code in this paper can be found at \url{https://github.com/f-blan/NOVO_Surrogate_Model} \citep{Blangiardi_ai_2026}.

\bibliographystyle{elsarticle-harv}
\bibliography{refs}

@Article{PHITS333_ref,
  author    = {Tatsuhiko Sato and Yosuke Iwamoto and Shintaro Hashimoto and Tatsuhiko Ogawa and Takuya Furuta and Shin-Ichiro Abe and Takeshi Kai and Yusuke Matsuya and Norihiro Matsuda and Yuho Hirata and Takuya Sekikawa and Lan Yao and Pi-En Tsai and Hunter N. Ratliff and Hiroshi Iwase and Yasuhito Sakaki and Kenta Sugihara and Nobuhiro Shigyo and Lembit Sihver and Koji Niita},
  journal   = {Journal of Nuclear Science and Technology},
  title     = {Recent improvements of the particle and heavy ion transport code system – {PHITS} version 3.33},
  year      = {2024},
  number    = {1},
  pages     = {127--135},
  volume    = {61},
  doi       = {10.1080/00223131.2023.2275736},
  publisher = {Taylor \& Francis},
}

@misc{gottschalk_techniques_2012,
	title = {Techniques of {Proton} {Radiotherapy}: {Transport} {Theory}},
	copyright = {arXiv.org perpetual, non-exclusive license},
	shorttitle = {Techniques of {Proton} {Radiotherapy}},
	doi = {10.48550/ARXIV.1204.4470},
	publisher = {arXiv},
	author = {Gottschalk, Bernard},
	year = {2012},
	note = {Version Number: 2},
}

@Article{JENDL5_ref,
  author    = {Iwamoto, Osamu and Iwamoto, Nobuyuki and Kunieda, Satoshi and Minato, Futoshi and Nakayama, Shinsuke and Abe, Yutaka and Tsubakihara, Kohsuke and Okumura, Shin and Ishizuka, Chikako and Yoshida, Tadashi and others},
  journal   = {Journal of Nuclear Science and Technology},
  title     = {{Japanese evaluated nuclear data library version 5: JENDL-5}},
  year      = {2023},
  number    = {1},
  pages     = {1--60},
  volume    = {60},
  doi       = {10.1080/00223131.2022.2141903},
  publisher = {Taylor \& Francis},
}

@inproceedings{emode_ref1,
  author  = {Niita, Koji and Iwamoto, Yosuke and Sato, Tatsuhiko and Iwase, Hiroshi and Matsuda, Norihiro and Sakamoto, Yukio and Nakashima, Hiroshi},
  booktitle = {International Conference on Nuclear Data for Science and Technology},
  title   = {A new treatment of radiation behaviour beyond one-body observables},
  year    = {2007},
  pages   = {1167-1169},
  doi     = {10.1051/ndata:07398},
}

@Article{Meric_2023,
  author    = {Ilker Meric and Enver Alagoz and Liv B. Hysing and Toni Kögler and Danny Lathouwers and William R. B. Lionheart and John Mattingly and Jasmina Obhodas and Guntram Pausch and Helge E. S. Pettersen and Hunter N. Ratliff and Marta Rovituso and Sonja M. Schellhammer and Lena M. Setterdahl and Kyrre Skjerdal and Edmond Sterpin and Davorin Sudac and Joseph A. Turko and Kristian S. Ytre-Hauge and},
  title     = {A hybrid multi-particle approach to range assessment-based treatment verification in particle therapy},
  journal   = {Scientific Reports},
  year      = {2023},
  volume    = {13},
  number    = {1},
  month     = {apr},
  pages     = {6709},
  doi       = {10.1038/s41598-023-33777-w},
  publisher = {Nature Publishing Group UK London},
}

@Article{Onizuka2016,
  author    = {Onizuka, Ryota and Araki, Fujio and Ohno, Takeshi and Nakaguchi, Yuji and Kai, Yudai and Tomiyama, Yuuki and Hioki, Kazunari},
  journal   = {Radiological physics and technology},
  title     = {Accuracy of dose calculation algorithms for virtual heterogeneous phantoms and intensity-modulated radiation therapy in the head and neck},
  year      = {2016},
  pages     = {77--87},
  volume    = {9},
  doi       = {10.1007/s12194-015-0336-z},
  publisher = {Springer},
}

@Article{INCL46_ref,
  author    = {Boudard, A. and Cugnon, J. and David, J.-C. and Leray, S. and Mancusi, D.},
  title     = {{New potentialities of the Li\`ege intranuclear cascade model for reactions induced by nucleons and light charged particles}},
  journal   = {Phys. Rev. C},
  year      = {2013},
  volume    = {87},
  issue     = {1},
  month     = {Jan},
  pages     = {014606},
  doi       = {10.1103/PhysRevC.87.014606},
  url       = {https://link.aps.org/doi/10.1103/PhysRevC.87.014606},
  numpages  = {28},
  publisher = {American Physical Society},
}

@Article{KUROTAMA_ref,
  author   = {Iida, Kei and Kohama, Akihisa and Oyamatsu, Kazuhiro},
  journal  = {Journal of the Physical Society of Japan},
  title    = {{Formula for Proton–Nucleus Reaction Cross Section at Intermediate Energies and Its Application}},
  year     = {2007},
  number   = {4},
  pages    = {044201},
  volume   = {76},
  doi      = {10.1143/JPSJ.76.044201},
}

@article{RTPHITS_ref1of2,
  author    = {Sato, Tatsuhiko and Furuta, Takuya and Liu, Yuwei and Naka, Sadahiro and Nagamori, Shushi and Kanai, Yoshikatsu and Watabe, Tadashi},
  journal   = {EJNMMI physics},
  title     = {Individual dosimetry system for targeted alpha therapy based on {PHITS} coupled with microdosimetric kinetic model},
  year      = {2021},
  pages     = {1--16},
  volume    = {8},
  doi       = {10.1186/s40658-020-00350-7},
  publisher = {Springer},
}

@Article{RTPHITS_ref2of2,
  author    = {Furuta, Takuya and Koba, Yusuke and Hashimoto, Shintaro and Chang, Weishan and Yonai, Shunsuke and Matsumoto, Shinnosuke and Ishikawa, Akihisa and Sato, Tatsuhiko},
  journal   = {Physics in Medicine \& Biology},
  title     = {Development of the {DICOM-based Monte Carlo} dose reconstruction system for a retrospective study on the secondary cancer risk in carbon ion radiotherapy},
  year      = {2022},
  month     = {jul},
  number    = {14},
  pages     = {145002},
  volume    = {67},
  doi       = {10.1088/1361-6560/ac7998},
  publisher = {IOP Publishing},
}

@Article{Clark2013_CTdata,
  author    = {Clark, Kenneth and Vendt, Bruce and Smith, Kirk and Freymann, John and Kirby, Justin and Koppel, Paul and Moore, Stephen and Phillips, Stanley and Maffitt, David and Pringle, Michael and others},
  journal   = {Journal of Digital Imaging},
  title     = {{The Cancer Imaging Archive (TCIA)}: maintaining and operating a public information repository},
  year      = {2013},
  pages     = {1045--1057},
  volume    = {26},
  doi       = {10.1007/s10278-013-9622-7},
  publisher = {Springer},
  url       = {https://doi.org/10.1007/s10278-013-9622-7},
}

@Misc{NIST_STAR_ref,
  author    = {Martin Berger and J. Coursey and M. Zucker and J. Chang},
  title     = {{ESTAR, PSTAR, and ASTAR: Computer Programs for Calculating Stopping-Power and Range Tables for Electrons, Protons, and Helium Ions (version 1.21)}},
  url       = {http://physics.nist.gov/Star},
  year      = {1999},
  language  = {en},
  month     = {1999-01-01},
  publisher = {http://physics.nist.gov/Star},
}

@article{parodi_vivo_2018,
	title = {\textit{{In} vivo} range verification in particle therapy},
	volume = {45},
	copyright = {http://onlinelibrary.wiley.com/termsAndConditions\#vor},
	issn = {0094-2405, 2473-4209},
	doi = {10.1002/mp.12960},
	language = {en},
	number = {11},
	journal = {Medical Physics},
	author = {Parodi, Katia and Polf, Jerimy C.},
	month = nov,
	year = {2018},
}

@article{mohan_proton_2017,
	title = {Proton therapy – {Present} and future},
	volume = {109},
	issn = {0169409X},
	doi = {10.1016/j.addr.2016.11.006},
	language = {en},
	journal = {Advanced Drug Delivery Reviews},
	author = {Mohan, Radhe and Grosshans, David},
	month = jan,
	year = {2017},
	pages = {26--44},
}

@article{moreno_intensity_2019,
	title = {Intensity modulated proton therapy ({IMPT}) – {The} future of {IMRT} for head and neck cancer},
	volume = {88},
	issn = {13688375},
	doi = {10.1016/j.oraloncology.2018.11.015},
	language = {en},
	journal = {Oral Oncology},
	author = {Moreno, Amy C. and Frank, Steven J. and Garden, Adam S. and Rosenthal, David I. and Fuller, Clifton D. and Gunn, Gary B. and Reddy, Jay P and Morrison, William H. and Williamson, Tyler D. and Holliday, Emma B. and Phan, Jack and Blanchard, Pierre},
	month = jan,
	year = {2019},
	pages = {66--74},
}

@article{farr_new_2018,
	title = {New horizons in particle therapy systems},
	volume = {45},
	issn = {0094-2405, 2473-4209},
	doi = {10.1002/mp.13193},
	language = {en},
	number = {11},
	journal = {Medical Physics},
	author = {Farr, Jonathan B. and Flanz, Jacob B. and Gerbershagen, Alexander and Moyers, Michael F.},
	month = nov,
	year = {2018},
}

@article{perali_prompt_2014,
	title = {Prompt gamma imaging of proton pencil beams at clinical dose rate},
	volume = {59},
	issn = {0031-9155, 1361-6560},
	doi = {10.1088/0031-9155/59/19/5849},
	language = {en},
	number = {19},
	journal = {Physics in Medicine and Biology},
	author = {Perali, I and Celani, A and Bombelli, L and Fiorini, C and Camera, F and Clementel, E and Henrotin, S and Janssens, G and Prieels, D and Roellinghoff, F and Smeets, J and Stichelbaut, F and Stappen, F Vander},
	month = oct,
	year = {2014},
	pages = {5849--5871},
}

@article{polf_imaging_2015,
	title = {Imaging of prompt gamma rays emitted during delivery of clinical proton beams with a {Compton} camera: feasibility studies for range verification},
	volume = {60},
	issn = {0031-9155, 1361-6560},
	shorttitle = {Imaging of prompt gamma rays emitted during delivery of clinical proton beams with a {Compton} camera},
	doi = {10.1088/0031-9155/60/18/7085},
	language = {en},
	number = {18},
	journal = {Physics in Medicine and Biology},
	author = {Polf, Jerimy C and Avery, Stephen and Mackin, Dennis S and Beddar, Sam},
	month = sep,
	year = {2015},
	pages = {7085--7099},
}

@article{zhang_calculation_2009,
	title = {Calculation of water equivalent thickness of materials of arbitrary density, elemental composition and thickness in proton beam irradiation},
	volume = {54},
	issn = {0031-9155, 1361-6560},
	doi = {10.1088/0031-9155/54/6/001},
	language = {en},
	number = {6},
	journal = {Physics in Medicine and Biology},
	author = {Zhang, Rui and Newhauser, Wayne D},
	month = mar,
	year = {2009},
	pages = {1383--1395},
}

@article{howell_secondary_2014,
	title = {Secondary neutron spectrum from 250‐{MeV} passively scattered proton therapy: {Measurement} with an extended‐range {Bonner} sphere system},
	volume = {41},
	copyright = {http://creativecommons.org/licenses/by/3.0/},
	issn = {0094-2405, 2473-4209},
	shorttitle = {Secondary neutron spectrum from 250‐{MeV} passively scattered proton therapy},
	doi = {10.1118/1.4892929},
	language = {en},
	number = {9},
	journal = {Medical Physics},
	author = {Howell, Rebecca M. and Burgett, E. A.},
	month = sep,
	year = {2014},
}

@article{hong_pencil_1996,
	title = {A pencil beam algorithm for proton dose calculations},
	volume = {41},
	issn = {0031-9155, 1361-6560},
	doi = {10.1088/0031-9155/41/8/005},
	language = {en},
	number = {8},
	journal = {Physics in Medicine and Biology},
	author = {Hong, Linda and Goitein, Michael and Bucciolini, Marta and Comiskey, Robert and Gottschalk, Bernard and Rosenthal, Skip and Serago, Chris and Urie, Marcia},
	month = aug,
	year = {1996},
	pages = {1305--1330},
}

@article{soukup_pencil_2005,
	title = {A pencil beam algorithm for intensity modulated proton therapy derived from {Monte} {Carlo} simulations},
	volume = {50},
	issn = {0031-9155, 1361-6560},
	doi = {10.1088/0031-9155/50/21/010},
	language = {en},
	number = {21},
	journal = {Physics in Medicine and Biology},
	author = {Soukup, Martin and Fippel, Matthias and Alber, Markus},
	month = nov,
	year = {2005},
	pages = {5089--5104},
}

@article{neishabouri_dose_2020,
	title = {Long short-term memory networks for proton dose calculation in highly heterogeneous tissues},
	volume = {48},
	issn = {4},
	doi = {https://doi.org/10.1002/mp.14658},
	language = {en},
	journal = {Medical Physics},
	author = {Neishabouri, Ahmad and Wahl, Niklas and Mairani, Andrea and Köthe, Ullrich and Bangert, Mark},
	month = dec,
	year = {2020},
	pages = {1893--1908},
}

@article{pastor-serrano_millisecond_2022,
	title = {Millisecond speed deep learning based proton dose calculation with {Monte} {Carlo} accuracy},
	volume = {67},
	issn = {0031-9155, 1361-6560},
	doi = {10.1088/1361-6560/ac692e},
	language = {en},
	number = {10},
	journal = {Physics in Medicine \& Biology},
	author = {Pastor-Serrano, Oscar and Perkó, Zoltán},
	month = may,
	year = {2022},
	pages = {105006},
}

@article{wu_improving_2021,
	title = {Improving proton dose calculation accuracy by using deep learning},
	volume = {2},
	issn = {2632-2153},
	doi = {10.1088/2632-2153/abb6d5},
	language = {en},
	number = {1},
	journal = {Machine Learning: Science and Technology},
	author = {Wu, Chao and Nguyen, Dan and Xing, Yixun and Montero, Ana Barragan and Schuemann, Jan and Shang, Haijiao and Pu, Yuehu and Jiang, Steve},
	month = mar,
	year = {2021},
	pages = {015017},
}

@article{marafini_mondo_2017,
	title = {{MONDO}: a neutron tracker for particle therapy secondary emission characterisation},
	volume = {62},
	issn = {0031-9155, 1361-6560},
	shorttitle = {{MONDO}},
	doi = {10.1088/1361-6560/aa623a},
	language = {en},
	number = {8},
	journal = {Physics in Medicine and Biology},
	author = {Marafini, M and Gasparini, L and Mirabelli, R and Pinci, D and Patera, V and Sciubba, A and Spiriti, E and Stoppa, D and Traini, G and Sarti, A},
	month = apr,
	year = {2017},
	pages = {3299--3312},
}

@article{lerendegui-marco_simultaneous_2022,
	title = {Simultaneous neutron and gamma imaging system for real time range and dose monitoring in {Hadron} {Therapy} and nuclear security applications},
	volume = {261},
	copyright = {https://creativecommons.org/licenses/by/4.0/},
	issn = {2100-014X},
	doi = {10.1051/epjconf/202226105001},
	language = {en},
	journal = {EPJ Web of Conferences},
	author = {Lerendegui-Marco, J. and Balibrea-Correa, J. and Babiano-Suárez, V. and Caballero, L. and Domingo-Pardo, C. and Ladarescu, I.},
	editor = {Mackova, A. and Lorenz, K. and Vantomme, A.},
	year = {2022},
	pages = {05001},
}

@article{anferov_analytic_2010,
	title = {Analytic estimates of secondary neutron dose in proton therapy},
	volume = {55},
	issn = {0031-9155, 1361-6560},
	doi = {10.1088/0031-9155/55/24/008},
	language = {en},
	number = {24},
	journal = {Physics in Medicine and Biology},
	author = {Anferov, V},
	month = dec,
	year = {2010},
	pages = {7509--7522},
}

@article{setterdahl_enhancing_2025,
	title = {Enhancing image quality in fast neutron-based range verification of proton therapy using a deep learning-based prior in {LM}-{MAP}-{EM} reconstruction},
	copyright = {https://iopscience.iop.org/page/copyright},
	issn = {0031-9155, 1361-6560},
	doi = {10.1088/1361-6560/ade198},
	language = {en},
	journal = {Physics in Medicine \& Biology},
	author = {Setterdahl, Lena Marie and Skjerdal, Kyrre and Ratliff, Hunter Nathaniel and Ytre-Hauge, Kristian Smeland and Lionheart, William R B and Holman, Sean and Pettersen, Helge Egil Seime and Blangiardi, Francesco and Lathouwers, Danny and Meric, Ilker},
	month = jun,
	year = {2025},
}

@article{setterdahl_evaluating_2025,
	title = {Evaluating impact of detector arrangement and position resolution effect on a fast neutron-based range verification system for proton therapy},
	volume = {234},
	issn = {0969806X},
	doi = {10.1016/j.radphyschem.2025.112793},
	language = {en},
	journal = {Radiation Physics and Chemistry},
	author = {Setterdahl, Lena M. and Lionheart, William R.B. and Lathouwers, Danny and Ratliff, Hunter N. and Skjerdal, Kyrre and Meric, Ilker},
	month = sep,
	year = {2025},
	pages = {112793},
}

@article{gueth_machine_2013,
	title = {Machine learning-based patient specific prompt-gamma dose monitoring in proton therapy},
	volume = {58},
	copyright = {http://iopscience.iop.org/info/page/text-and-data-mining},
	issn = {0031-9155, 1361-6560},
	doi = {10.1088/0031-9155/58/13/4563},
	language = {en},
	number = {13},
	journal = {Physics in Medicine and Biology},
	author = {Gueth, P and Dauvergne, D and Freud, N and Létang, J M and Ray, C and Testa, E and Sarrut, D},
	month = jul,
	year = {2013},
	pages = {4563--4577},
}

@article{paganetti_range_2012,
	title = {Range uncertainties in proton therapy and the role of {Monte} {Carlo} simulations},
	volume = {57},
	issn = {0031-9155, 1361-6560},
	doi = {10.1088/0031-9155/57/11/R99},
	language = {en},
	number = {11},
	journal = {Physics in Medicine and Biology},
	author = {Paganetti, Harald},
	month = jun,
	year = {2012},
	pages = {R99--R117},
}

@article{burlacu_deterministic_2023,
	title = {A {Deterministic} {Adjoint}-{Based} {Semi}-{Analytical} {Algorithm} for {Fast} {Response} {Change} {Computations} in {Proton} {Therapy}},
	volume = {52},
	issn = {2332-4309, 2332-4325},
	doi = {10.1080/23324309.2023.2166077},
	language = {en},
	number = {1},
	journal = {Journal of Computational and Theoretical Transport},
	author = {Burlacu, Tiberiu and Lathouwers, Danny and Perkó, Zoltán},
	month = jan,
	year = {2023},
	pages = {1--41},
}

@article{ashby_efficient_2025,
	title = {Efficient proton transport modelling for proton beam therapy and biological quantification},
	volume = {90},
	issn = {0303-6812, 1432-1416},
	doi = {10.1007/s00285-025-02212-1},
	language = {en},
	number = {5},
	journal = {Journal of Mathematical Biology},
	author = {Ashby, Ben S. and Chronholm, Veronika and Hajnal, Daniel K. and Lukyanov, Alex and MacKenzie, Katherine and Pim, Aaron and Pryer, Tristan},
	month = may,
	year = {2025},
	pages = {47},
}

@misc{zhang_deterministic_2025,
	title = {A deterministic solver for the linear {Boltzmann} model of a single mono-directional proton beam},
	doi = {10.48550/arXiv.2504.00340},
	language = {en},
	publisher = {arXiv},
	author = {Zhang, Xiaojiang and Bai, Xuemin and Tang, Min},
	month = apr,
	year = {2025},
}

@inproceedings{stammer_deterministic_2025,
	title = {A {Deterministic} {Dynamical} {Low}-rank {Approach} for {Charged} {Particle} {Transport}},
	doi = {10.13182/xyz-46813},
	language = {en},
	booktitle = {International Conference on Mathematics and Computational Methods Applied to Nuclear Science and Engineering},
	author = {Stammer, Pia and Burlacu, Tiberiu and Wahl, Niklas and Lathouwers, Danny and Kusch, Jonas},
	month = jan,
	year = {2025},
    pages = {556-565},
}

@article{schneider_neutrons_2017,
	title = {Neutrons in active proton therapy: {Parameterization} of dose and dose equivalent},
	volume = {27},
	copyright = {https://www.elsevier.com/tdm/userlicense/1.0/},
	issn = {09393889},
	shorttitle = {Neutrons in active proton therapy},
	doi = {10.1016/j.zemedi.2016.07.001},
	language = {en},
	number = {2},
	journal = {Zeitschrift für Medizinische Physik},
	author = {Schneider, Uwe and Hälg, Roger A. and Lomax, Tony},
	month = jun,
	year = {2017},
	pages = {113--123},
}

@article{xiao_prompt_2024,
	title = {Prompt gamma emission prediction using a long short-term memory network},
	volume = {69},
	issn = {0031-9155, 1361-6560},
	doi = {10.1088/1361-6560/ad8e2a},
	language = {en},
	number = {23},
	journal = {Physics in Medicine \& Biology},
	author = {Xiao, Fan and Radonic, Domagoj and Kriechbaum, Michael and Wahl, Niklas and Neishabouri, Ahmad and Delopoulos, Nikolaos and Parodi, Katia and Corradini, Stefanie and Belka, Claus and Kurz, Christopher and Landry, Guillaume and Dedes, George},
	month = dec,
	year = {2024},
	pages = {235003},
}

@misc{kovachki_neural_2024,
	title = {Neural {Operator}: {Learning} {Maps} {Between} {Function} {Spaces}},
	shorttitle = {Neural {Operator}},
	doi = {10.5555/3648699.3648788},
	language = {en},
	author = {Kovachki, Nikola and Li, Zongyi and Liu, Burigede and Azizzadenesheli, Kamyar and Bhattacharya, Kaushik and Stuart, Andrew and Anandkumar, Anima},
	month = may,
	year = {2024},
	note = {arXiv:2108.08481 [cs, math]},
}

@article{azizzadenesheli_neural_2024,
	title = {Neural operators for accelerating scientific simulations and design},
	volume = {6},
	issn = {2522-5820},
	doi = {10.1038/s42254-024-00712-5},
	language = {en},
	number = {5},
	journal = {Nature Reviews Physics},
	author = {Azizzadenesheli, Kamyar and Kovachki, Nikola and Li, Zongyi and Liu-Schiaffini, Miguel and Kossaifi, Jean and Anandkumar, Anima},
	month = apr,
	year = {2024},
	pages = {320--328},
}

@misc{li_fourier_2021,
	title = {Fourier {Neural} {Operator} for {Parametric} {Partial} {Differential} {Equations}},
	doi = {10.48550/arXiv.2010.08895},
	language = {en},
	publisher = {arXiv},
	author = {Li, Zongyi and Kovachki, Nikola and Azizzadenesheli, Kamyar and Liu, Burigede and Bhattacharya, Kaushik and Stuart, Andrew and Anandkumar, Anima},
	month = may,
	year = {2021},
}

@article{hu_predicting_2025,
	title = {Predicting radiation distribution via neural operator trained on basis function-generated data},
	volume = {315},
	issn = {00104655},
	doi = {10.1016/j.cpc.2025.109710},
	language = {en},
	journal = {Computer Physics Communications},
	author = {Hu, Ankang and Li, Kaiwen and Qiu, Rui and Li, Junli},
	month = oct,
	year = {2025},
	pages = {109710},
}

@Article{Tjelta2023,
  author    = {Tjelta, Johannes and Fjæra, Lars Fredrik and Ytre-Hauge, Kristian Smeland and Boer, Camilla Grindeland and Stokkevåg, Camilla Hanquist},
  journal   = {Physics in Medicine \& Biology},
  title     = {{A systematic approach for calibrating a Monte Carlo code to a treatment planning system for obtaining dose, LET, variable proton RBE and out-of-field dose}},
  year      = {2023},
  month     = {nov},
  number    = {22},
  pages     = {225010},
  volume    = {68},
  doi       = {10.1088/1361-6560/ad0281},
  publisher = {IOP Publishing},
}

@inproceedings{emode_ref2,
  author  = {Iwamoto, Yosuke and Niita, Koji and Sakamoto, Yukio and Sato, Tatsuhiko and Matsuda, Norihiro},
  booktitle = {International Conference on Nuclear Data for Science and Technology},
  title   = {Validation of the event generator mode in the {PHITS} code and its application},
  year    = {2007},
  pages   = {945-948},
  doi     = {10.1051/ndata:07417},
}

@article{emode_ref3,
  author  = {Iwamoto, Yosuke and Niita, Koji and Sato, Tatsuhiko and Matsuda, Norihiro and Iwase, Hiroshi and Nakashima, Hiroshi and Sakamoto, Yukio},
  journal = {Prog Nucl Sci Technol},
  title   = {Application and validation of event generator in the {PHITS} code for the low-energy neutron-induced reactions},
  year    = {2011},
  pages   = {931-935},
  volume  = {2},
  doi = {10.15669/pnst.2.931}
}

@misc{kossaifi2024neural,
   title={A Library for Learning Neural Operators},
   author={Jean Kossaifi and Nikola Kovachki and
   Zongyi Li and David Pitt and
   Miguel Liu-Schiaffini and Robert Joseph George and
   Boris Bonev and Kamyar Azizzadenesheli and
   Julius Berner and Anima Anandkumar},
   year={2024},
   eprint={2412.10354},
   archivePrefix={arXiv},
   primaryClass={cs.LG}
}

@inproceedings{Chen_2016, series={KDD ’16},
   title={XGBoost: A Scalable Tree Boosting System},
   url={http://dx.doi.org/10.1145/2939672.2939785},
   DOI={10.1145/2939672.2939785},
   booktitle={Proceedings of the 22nd ACM SIGKDD International Conference on Knowledge Discovery and Data Mining},
   publisher={ACM},
   author={Chen, Tianqi and Guestrin, Carlos},
   year={2016},
   month=aug, pages={785–794},
   collection={KDD ’16} }

@inproceedings{feydy2019interpolating,
    title={Interpolating between Optimal Transport and MMD using Sinkhorn Divergences},
    author={Feydy, Jean and S{\'e}journ{\'e}, Thibault and Vialard, Fran{\c{c}}ois-Xavier and Amari, Shun-ichi and Trouve, Alain and Peyr{\'e}, Gabriel},
    booktitle={The 22nd International Conference on Artificial Intelligence and Statistics},
    pages={2681-2690},
    year={2019}
}

@article{Biggs2022, doi = {10.21105/joss.04555}, year = {2022}, publisher = {The Open Journal}, volume = {7}, number = {78}, pages = {4555}, author = {Biggs, Simon and Jennings, Matthew and Swerdloff, Stuart and Chlap, Phillip and Lane, Derek and Rembish, Jacob and McAloney, Jacob and King, Paul and Ayala, Rafael and Guan, Fada and Lambri, Nicola and Crewson, Cody and Sobolewski, Matthew}, title = {PyMedPhys: A community effort to develop an open, Python-based standard library for medical physics applications}, journal = {Journal of Open Source Software} }

@article{Ratliff2025_PHITSTools, 
  doi = {10.21105/joss.08311}, 
  year = {2025}, 
  publisher = {The Open Journal}, 
  volume = {10}, 
  number = {113}, 
  pages = {8311}, 
  author = {Ratliff, Hunter N.}, 
  title = {{The PHITS Tools Python package for parsing, organizing, and analyzing results from the PHITS radiation transport and DCHAIN activation codes}}, 
  journal = {Journal of Open Source Software} 
}

@article{
    gao2025dynamic,
    title={Dynamic Schwartz-Fourier Neural Operator for Enhanced Expressive Power},
    author={Wenhan Gao and Jian Luo and Ruichen Xu and Yi Liu},
    journal={Transactions on Machine Learning Research},
    year={2025},
    url={https://openreview.net/forum?id=B0E2yjrNb8},
}

@misc{li_fourier_2024,
	title = {Fourier {Neural} {Operator} with {Learned} {Deformations} for {PDEs} on {General} {Geometries}},
	doi = {10.5555/3648699.3649087},
	language = {en},
	author = {Li, Zongyi and Huang, Daniel Zhengyu and Liu, Burigede and Anandkumar, Anima},
	month = may,
	year = {2024},
}

@misc{Ratliff_Rodare,
  author       = {Ratliff, Hunter N. and
                  Blangiardi, Francesco},
  title        = {{PHITS simulations of neutron and gamma-ray production from and transport of 70--250 MeV protons in heterogeneous 1D tissue phantoms}},
  year         = 2026,
  note         = {{Rodare}, {\textit{(to be published in February-March 2026)}}},
  doi          = {10.14278/rodare.3997},
}

@misc{
Blangiardi_ai_2026,
  author  = {Blangiardi, Francesco},
  title   = {NOVO surrogate model},
  year     = 2026,
  version = {v1.0.0},
  date    = {2026-01-27},
  url     = {https://github.com/f-blan/NOVO_Surrogate_Model},
}

@misc{Blangiardi_Rodare,
  author       = {Blangiardi, Francesco and Ratliff, Hunter N.},
  title        = {Proton and Neutron reduced phase space for surrogate modeling of Proton Therapy from PHITS simulations},
  year         = {2025},
  doi          = {10.14278/rodare.4127},
}

@inproceedings{Setterdahl_unets_2024,
    doi = {10.1007/978-3-031-66955-2_16},
    author={Setterdahl, Lena M. and Lionheart, William R. B. and Holman, Sean and Skjerdal, Kyrre and Ratliff, Hunter N. and Ytre-Hauge, Kristian Smeland and Lathouwers, Danny and Meric, Ilker},
    title={Image Reconstruction for Proton Therapy Range Verification via U-NETs},
    booktitle={Yap, M.H., Kendrick, C., Behera, A., Cootes, T., Zwiggelaar, R. (eds). Medical Image Understanding and Analysis},
    year={2024},
    pages={232--244},
}

@article{shrestha_stray_2022,
	title = {Stray neutron radiation exposures from proton therapy: physics-based analytical models of neutron spectral fluence, kerma and absorbed dose},
	volume = {67},
	issn = {0031-9155, 1361-6560},
	shorttitle = {Stray neutron radiation exposures from proton therapy},
	doi = {10.1088/1361-6560/ac7377},
	language = {en},
	number = {12},
	journal = {Physics in Medicine \& Biology},
	author = {Shrestha, Suman and Newhauser, Wayne D and Donahue, William P and Pérez-Andújar, Angélica},
	month = jun,
	year = {2022},
	pages = {125019},
}

@article{taylor_pencil_2017,
	title = {Pencil {Beam} {Algorithms} {Are} {Unsuitable} for {Proton} {Dose} {Calculations} in {Lung}},
	volume = {99},
	issn = {03603016},
	doi = {10.1016/j.ijrobp.2017.06.003},
	language = {en},
	number = {3},
	journal = {International Journal of Radiation Oncology*Biology*Physics},
	author = {Taylor, Paige A. and Kry, Stephen F. and Followill, David S.},
	month = nov,
	year = {2017},
	pages = {750--756},
}

@article{ytre-hauge_monte_2019,
	title = {A {Monte} {Carlo} feasibility study for neutron based real-time range verification in proton therapy},
	volume = {9},
	issn = {2045-2322},
	doi = {10.1038/s41598-019-38611-w},
	language = {en},
	number = {1},
	journal = {Scientific Reports},
	author = {Ytre-Hauge, Kristian Smeland and Skjerdal, Kyrre and Mattingly, John and Meric, Ilker},
	month = feb,
	year = {2019},
}

@article{dhakal_symmetric_2014,
	title = {A symmetric probabilistic \textit{y} -index for {Monte} {Carlo} dose comparisons},
	volume = {59},
	copyright = {http://iopscience.iop.org/info/page/text-and-data-mining},
	issn = {0031-9155, 1361-6560},
	doi = {10.1088/0031-9155/59/16/N153},
	language = {en},
	number = {16},
	journal = {Physics in Medicine and Biology},
	author = {Dhakal, Tilak R and Yepes, Pablo},
	month = aug,
	year = {2014},
	pages = {153--161},
}

@article{kusch_robust_2023,
	title = {A robust collision source method for rank adaptive dynamical low-rank approximation in radiation therapy},
	volume = {57},
	copyright = {https://creativecommons.org/licenses/by/4.0},
	issn = {2822-7840, 2804-7214},
	doi = {10.1051/m2an/2022090},
	language = {en},
	number = {2},
	journal = {ESAIM: Mathematical Modelling and Numerical Analysis},
	author = {Kusch, Jonas and Stammer, Pia},
	month = mar,
	year = {2023},
	pages = {865--891},
}

@article{pinto_filtering_2020,
	title = {A filtering approach for {PET} and {PG} predictions in a proton treatment planning system},
	volume = {65},
	issn = {0031-9155, 1361-6560},
	doi = {10.1088/1361-6560/ab8146},
	language = {en},
	number = {9},
	journal = {Physics in Medicine \& Biology},
	author = {Pinto, M and Kröniger, K and Bauer, J and Nilsson, R and Traneus, E and Parodi, K},
	month = may,
	year = {2020},
	pages = {095014},
}

\end{document}